\documentclass[twocolumn]{aastex62}
\usepackage{amsbsy}
\usepackage{amsfonts}
\usepackage{amssymb}
\usepackage{bm,ulem}
\usepackage{mathrsfs}
\usepackage{pifont}
\usepackage{stmaryrd}
\usepackage{textcomp,bookmark}
\usepackage{xspace}
\usepackage{amsmath,amsxtra}
\usepackage[OT2,OT1]{fontenc}
\usepackage{graphicx}
\usepackage{hyperref}
\usepackage{natbib}
\usepackage{array,colortbl,diagbox}
\usepackage{harpoon}
\usepackage{ulem}

\newcommand{\msun}{\ifmmode {{\rm M}_\odot} \else  {${\rm M}_\odot$} \fi}
\newcommand{\hms}{\ifmmode {h^{-1}{\rm M}_\odot} \else  {$h^{-1}{\rm M}_\odot$} \fi}
\newcommand{\mpc}{\ifmmode {{\rm Mpc}} \else  {${\rm Mpc}$} \fi}
\newcommand{\kpc}{\ifmmode {{\rm kpc}} \else  {${\rm kpc}$} \fi}
\newcommand{\hmpc}{\ifmmode {h^{-1}{\rm Mpc}} \else  {$h^{-1}{\rm Mpc}$} \fi}
\newcommand{\hkpc}{\ifmmode {h^{-1}{\rm kpc}} \else  {$h^{-1}{\rm kpc}$} \fi}
\newcommand{\amin}{\ifmmode {{\rm arcmin}} \else  {${\rm arcmin}$} \fi}
\newcommand{\mv}{\ifmmode {{M}_{\rm vir}} \else  {${M}_{\rm vir}$} \fi}
\newcommand{\mt}{\ifmmode {{M}_{200}} \else  {${M}_{200}$} \fi}
\newcommand{\mlow}{\ifmmode {{M}_{\rm low}} \else  {${M}_{\rm low}$} \fi}
\newcommand{\neff}{\ifmmode {{n}_{\rm eff}} \else  {${n}_{\rm eff}$} \fi}
\newcommand{\cv}{\ifmmode {{c}_{\rm vir}} \else  {${c}_{\rm vir}$} \fi}
\newcommand{\ct}{\ifmmode {{c}_{200}} \else  {${c}_{200}$} \fi}
\newcommand{\rv}{\ifmmode {{r}_{\rm vir}} \else  {${r}_{\rm vir}$} \fi}
\newcommand{\rt}{\ifmmode {{r}_{200}} \else  {${r}_{200}$} \fi}
\newcommand{\rdlt}{\ifmmode {{r}_{\scriptscriptstyle\Delta}} \else  {${r}_{\scriptscriptstyle\Delta}$} \fi}
\newcommand{\mdlt}{\ifmmode {{M}_{\scriptscriptstyle\Delta}} \else  {${M}_{\scriptscriptstyle\Delta}$} \fi}
\newcommand{\cdlt}{\ifmmode {{c}_{\scriptscriptstyle\Delta}} \else  {${c}_{\scriptscriptstyle\Delta}$} \fi}
\newcommand{\smM}{\ifmmode {\scriptscriptstyle M} \else  {${\scriptscriptstyle M}$} \fi}
\newcommand{\smT}{\ifmmode {\scriptscriptstyle T} \else  {${\scriptscriptstyle T}$} \fi}
\newcommand{\zbpz}{\ifmmode {z_{\scriptscriptstyle BPZ}} \else  {${z_{\scriptscriptstyle BPZ}}$} \fi}
\newcommand{\hh}{\ifmmode {h^{-1}} \else  {$h^{-1}$} \fi}
\newcommand{\vth}{\ifmmode {\vec{\theta}} \else  {$\vec{\theta}$} \fi}
\def\etal{et al.}
\def\ie{{\frenchspacing\it i.e.}}
\def\eg{{\frenchspacing\it e.g.}}

\newcommand{\remove}[1]{}
\newcommand{\yppn}{\ifmmode {{\gamma}_{\scriptscriptstyle {\rm PPN}}} \else  {${\gamma}_{\scriptscriptstyle {\rm PPN}}$} \fi}
\newcommand{\rpr}{\ifmmode {r^\prime} \else  {$r^\prime$} \fi}

\newcommand{\lgg}{\ifmmode {{\rm log}} \else  {${\rm log}$} \fi}
\newcommand{\sn}{\ifmmode {\sigma_n} \else  {$\sigma_n$} \fi}
\newcommand{\snsq}{\ifmmode {\sigma_n^2} \else  {$\sigma_n^2$} \fi}

\newcommand{\dd}{{\rm d}}

\begin{document}
\title{An accurate analytic mass model for lensing galaxies}

\author[0000-0001-9781-6863]{Wei Du}\thanks{E-mail: \url{duwei@bao.ac.cn}}
\affiliation{National Astronomical Observatories, Chinese Academy of Science, Beijing, 100101, P.R.China}

\author[0000-0003-4726-6714]{Gong-Bo Zhao}\thanks{E-mail: \url{gbzhao@nao.cas.cn}}
\affiliation{National Astronomical Observatories, Chinese Academy of Science, Beijing, 100101, P.R.China}
\affiliation{School of Astronomy and Space Science, University of Chinese Academy of Sciences, Beijing, 100049, P.R.China}

\author{Zuhui Fan}
\affiliation{South-Western Institute for Astronomy Research, Yunnan University, Kunming, 650500, P.R.China}
\affiliation{Department of Astronomy, Peking University, Beijing, 100871, P.R.China}

\author[0000-0002-9063-698X]{Yiping Shu}
\affiliation{Institute of Astronomy, University of Cambridge, Madingley Road, Cambridge, CB3 0HA, UK}

\author{Ran Li}
\affiliation{National Astronomical Observatories, Chinese Academy of Science, Beijing, 100101, P.R.China}

\author[0000-0001-8317-2788]{Shude Mao}
\affiliation{Department of Astronomy, Tsinghua University, Beijing, 100084, P.R.China}
\affiliation{National Astronomical Observatories, Chinese Academy of Science, Beijing, 100101, P.R.China}

\begin{abstract}
  We develop an analytic mass model for lensing galaxies, based on a broken power-law (BPL) density profile, which is a power-law profile with a mass deficit or surplus in the central region. Under the assumption of an elliptically symmetric surface mass distribution, the deflection angle and magnification can be evaluated analytically for this new model. We compute the theoretical prediction for various quantities, including the volume and surface mass density profiles of the galaxies, and the aperture and luminosity-weighted line-of-sight velocity dispersions, and compare them to those measured from the Illustris simulation. We find an excellent agreement between our model prediction and the simulation, which validates our modeling. The high efficiency and accuracy of our model manifests itself as a promising tool for studying properties of galaxies with strong lensing.
\end{abstract}


\section{Introduction}
The phenomenon of gravitational lensing is caused by light bending in spacetime and is thus sensitive to the geometry of the universe and the matter distribution therein. The relevant observables are the light magnifications (or demagnifications), position displacements, shape distortions, time delays, and so on \citep{SWM2006book}. Since the first discovery of the strong gravitational lens Q0957+561 \citep{1979Natur.279..381W}, gravitational lensing (including micro, weak, and strong) has become one of the most powerful techniques to address the issues on a wide range of scales, \eg~from the scales of planets, galaxies, and galaxy clusters, to cosmic scales \citep{2006PhR...429....1L,2010CQGra..27w3001B,2010ARA&A..48...87T,2012ARA&A..50..411G,2012RAA....12..947M,2013SSRv..177...75H,2015RPPh...78h6901K,2017SchpJ..1232440B}.

On galaxy scales, strong lensing (SL) has been studied extensively. The multiple images or extended arcs are widely used to constrain the mass distribution of galaxies \citep{2006ApJ...649..599K,2010ApJ...724..511A,2010GReGr..42.2151K,2012ApJ...747L..15G,2017MNRAS.464.4823B}. The time delays between multiply imaged quasars provide an approach to constrain the Hubble constant \citep{2010ApJ...711..201S,2013ApJ...766...70S,2017MNRAS.465.4914B,2017MNRAS.465.4895W,2018MNRAS.474.4648S}. The statistics of giant arcs have also been investigated to constrain cosmology \citep{2013SSRv..177...31M}. Combined with stellar dynamics, galactic models and gravity theories can also be tested \citep{2006PhRvD..74f1501B,2010ApJ...708..750S,2018Sci...360.1342C}. The lensing mass distribution is essential for almost all SL-related studies. However, until now, there have been no generic lensing mass models proposed in observations or simulations for SL analyses. The reasons may be as follows:

(I) Observationally, SL images are vulnerable to the point-spread function, image pixelization, and light contamination from foreground lenses \citep{2015A&A...577A..85B}. Moreover, the lensing patterns may differ in different colors if the shape of the background galaxies is sensitive to the observational waveband \citep{2006ApJ...638..703B,2007ApJ...671.1196M}, although the SL effect itself is color independent. These observational uncertainties inevitably complicate SL analyses and make it hard to accurately measure the mass distributions of galaxies.

(II) Although numerical simulations can help us find a universal mass density profile, \eg, the Navarro--Frenk--White profile \citep[NFW;][]{1997ApJ...490..493N}, for massive dark matter halos, the simulations have limitations, especially in central regions. Both the complexity of baryonic physics and the limitation of numerical resolutions can lead to unrealistic density profiles. For example, the cooling tends to steepen the density profiles while the feedback has an opposite effect. In addition, the gravitational softening can smooth the mass distribution of {\it N}-body systems in simulations \citep{2012MNRAS.425.1104B}.

(III) Degeneracies exist in the modeling of the lensing mass.
For example, due to the so-called source position transformation (SPT), different lensing mass models can have nearly the same lensed images \citep{2013A&A...559A..37S,2014A&A...564A.103S,2017MNRAS.464.4823B}. The complex degeneracies indicate that the parameter space may be full of local maxima. Strong priors should be added to shrink the parameter space to give better constraints on the lensing mass distributions.

(IV) In addition to the problems mentioned above, another challenge is to calculate the deflection field of a realistic lensing mass distribution, which is in general not spherical. In principle, this should not be a big issue because the deflection angles can always be evaluated using numerical integrals for a mass distribution. However, in practice, numerical integrals are computationally too expensive to be feasible for large samples.

So, how can we alleviate these problems?
If we put aside the observational effects mentioned above, we can notice that the main obstacle is to find a more realistic lensing mass model that allows for analytical calculation of the deflection angles.

As we know, actual galaxies show different morphologies, from irregular to regular (disk-like or elliptical) shapes.
Irregular galaxies present complex structures for which the mass distributions are hard to model. For regular galaxies, the triaxiality has been demonstrated by observations and {\it N}-body simulations.
Furthermore, it is found that the triaxiality of the isodensity surfaces may vary with radius. However, to the first-order approximation, we can assume that the triaxial mass distribution is homoeoidal, which means that the projected surface mass distribution is still homoeoidal, \ie~ elliptically symmetric \citep{1984MNRAS.208..511B,1990A&A...231...19S}.

\citet{1973ApJ...185..747B} and \citet[hereafter BK75]{1975ApJ...195...13B} first introduced the complex formulation of the deflection angles for mass distributions with a homoeoidal symmetry (see also \citealt{1984MNRAS.208..511B} for minor corrections). The equivalent formulations in real notation were given by \citet[hereafter S90]{1990A&A...231...19S} for the purpose of calculating the two components of deflection angles separately. Although these formulae are elegant with one-dimensional integrals, the analytical deflection angles have been derived only for a few lensing mass models.

Among the elliptical mass distributions, the widely investigated one is the softened power-law (SFPL) density profile, for which the surface mass distribution can be described by $\kappa_{\rm SFPL}\propto (R_{\rm el}^2+R_{\rm c}^2)^{-\nu/2}$, where $R_{\rm el}$ and $R_{\rm c}$ are the elliptical radius and the core radius, respectively. Based on the complex deflection formulation of BK75, \citet{1993ApJ...417..450K} found that analytical solutions exist for special cases of the SFPL model, with $\nu$ being integers. For $\nu=1$, it is the softened isothermal ellipsoid (SFIE) model \citep{1994A&A...284..285K,1998ApJ...495..157K}. The singular isothermal ellipsoid (SIE) model, which is a special case of the SFIE model with $R_{\rm c}=0$, is commonly used in SL-related studies.

Efforts were taken to derive the analytical deflections of SFPL model with an arbitrary slope $\nu$. Also adopting the BK75 formulation, \citet{1996ApJ...464...92G} presented the complex expression of the deflection angle for the singular power-law (SPL) profile, which was further investigated in detail in \citet{2015A&A...580A..79T}. By changing the variables in the formulation of S90 and using the polynomial expansions of the relevant integrand, \citet{1998ApJ...502..531B} found that the deflection angles for the flexible SFPL model can be written as a series and double series.

In addition to the formulae introduced by BK75 and S90, other strategies were also proposed to estimate the deflection angles of elliptical mass distributions. For instance, methods resorting to the Fourier series were investigated by \citet{1991A&A...247..269S} and \citet{1998ApJ...506...80C}, where the SFPL model was also inspected as a special case. \citet{2002ApJ...568..500C} also applied their Fourier series method to the general cusped two-power-law ellipsoidal profile.
All their results showed the necessity of double or even triple sums to calculate the deflection angles, which cannot be evaluated efficiently in most cases.

In light of the fact that the deflection angle can be expressed as a convolution product between the convergence $\kappa(\vec{x})$ and the kernel $\vec{x}/|\vec{x}|^2$, \citet{2014MNRAS.437.1051W} first deduced the analytical deflection angles for the flexible SFPL model using the Fourier transform. We now realize that there are obvious difficulties in calculating the deflection angles of elliptical mass distributions, even for the seemingly simple SFPL model.

Although the analytical deflection angles have already been derived for the more general SPL or SFPL models, these models have not yet been commonly used in SL analyses. One of the main reasons may be that their fidelities to the actual mass distributions have not been well tested. It is thus necessary to find a lensing mass model that allows for the analytical calculation of deflection angles and is realistic.

In this paper, we propose the broken power-law (BPL) profile for which both the deflection angles and magnifications can be analytically calculated. We find that the BPL model can well describe the surface mass distributions of galaxies in the Illustris simulation \citep{2014MNRAS.444.1518V,2015A&C....13...12N}. We also investigate the line-of-sight velocity dispersions (LOSVDs) for this new model.

The rest of the paper is organized as follows. In Section 2, we present basic formulations for the BPL model. Section 3 describes the simulated galaxies used to test the BPL model. Methods of density profile fittings are described in Section 4. The model fitting results are presented in Section 5, and the last section is devoted to the conclusion and discussions. More details can be found in the Appendix.

\section{The BPL model}
The BPL profile proposed in this paper has the following volume density profile,
\begin{equation}\label{eq:rho}
\rho(r)=\left\{
           \begin{array}{ll}
             \displaystyle\rho_c\left(r/r_c\right)^{-\alpha_c} &
\hbox{if $r\le r_c$} \\\\
             \displaystyle\rho_c\left(r/r_c\right)^{-\alpha}
& \hbox{if $r\ge r_c$} ,
           \end{array}
         \right.
\end{equation}
where $0\le\alpha_c<3$, $1<\alpha<3$, $r_c$ denotes the break radius, \ie~ the size of the central region for which the slope differs from the outer part, and $\rho_c$ is the density at $r_c$.

The total mass within $r$ is given by
\begin{equation}\label{eq:mass3d}
M(r)=\left\{
           \begin{array}{ll}
             \displaystyle\frac{4\pi\rho_c}{3-\alpha_c}r_c^{\alpha_c}r^{3-\alpha_c} &
\hbox{if $r\le r_c$} \\\\
             \displaystyle\frac{4\pi\rho_c}{3-\alpha}r_c^{\alpha}r^{3-\alpha}+m_0
& \hbox{if $r\ge r_c$}
           \end{array}
         \right.
\end{equation}
with
\begin{equation}\label{eq:mass0}
m_0=-\frac{4\pi\rho_c}{3-\alpha}\frac{\alpha-\alpha_c}{3-\alpha_c}r_c^3
\end{equation}
indicating a mass deficit or a surplus for $\alpha_c<\alpha$ or $\alpha_c>\alpha$, respectively, in the central region.

\subsection{The Projected Surface Mass Distribution}
By integrating the volume density profile $\rho(r)$ along the line of sight, which is taken to be the $Z$ direction here, we obtain the surface density profile,
\begin{eqnarray}\label{eq:surmass}
&&\Sigma(R)=2\int_0^\infty \rho(r)\dd Z= \nonumber\\
&&\left\{
           \begin{array}{ll}
             \displaystyle \mathcal{B}(\alpha)\rho_cr_c^\alpha R^{1-\alpha}-2\rho_cr_c\times & \\\\
             \displaystyle \tilde{z}\left[F\left(\frac{\alpha}{2},1;\frac{3}{2};\tilde{z}^2\right)-F\left(\frac{\alpha_c}{2},1;\frac{3}{2};\tilde{z}^2\right)\right] &
\hbox{if $R\le r_c$} \\\\
             \displaystyle \mathcal{B}(\alpha)\rho_cr_c^\alpha R^{1-\alpha}
& \hbox{if $R\ge r_c$},
           \end{array}
         \right. \nonumber\\
\end{eqnarray}
where $r^2=R^2+Z^2$, $F()$ is the Gauss hypergeometric function, $\tilde{z}=\sqrt{1-R^2/r_c^2}$ and
\begin{equation}
 \mathcal{B}(\alpha)={\rm Beta}\left(\frac{1}{2},\frac{\alpha-1}{2}\right)
 =\sqrt{\pi}\frac{\Gamma(\frac{\alpha-1}{2})}{\Gamma(\frac{\alpha}{2})}, 
\end{equation}
where $\Gamma(x)$ and $\rm Beta(x,y)$ are the complete gamma function and the beta function, respectively. The total mass within the projected radius $R$ is then
\begin{eqnarray}\label{eq:mass2d}
&&M_{\rm 2D}(R)=2\pi\int_0^R \Sigma(R)R \ {\rm d}R= \nonumber\\
&&\left\{
           \begin{array}{ll}
             \displaystyle 2\pi\frac{\mathcal{B}(\alpha)}{3-\alpha} \rho_cr_c^\alpha R^{3-\alpha}+m_0+\frac{4\pi \rho_c }{3}r_c^3\times & \\\\
             \displaystyle \tilde{z}^3\left[F\left(\frac{\alpha}{2},1;\frac{5}{2};\tilde{z}^2\right)-F\left(\frac{\alpha_c}{2},1;\frac{5}{2};\tilde{z}^2\right)\right] &
\hbox{if $R\le r_c$} \\\\
             \displaystyle 2\pi\frac{\mathcal{B}(\alpha)}{3-\alpha} \rho_cr_c^\alpha R^{3-\alpha}+m_0
& \hbox{if $R\ge r_c$}.
           \end{array}
         \right. \nonumber\\
\end{eqnarray}

In lensing analyses, what we care about is the convergence, which is the surface mass density scaled by the critical surface mass density,
\begin{equation}\label{eq:sigc}
\Sigma_{\rm crit}=\frac{c^2}{4\pi G}\frac{D_s}{D_dD_{ds}},
\end{equation}
where $D_s$, $D_d$, and $D_{ds}$ are the angular diameter distances from the observer to the background source and to the lens, and from the lens to the source, respectively. By further introducing a scale radius $b$, \ie~
\begin{equation}\label{eq:brhoc}
b^{\alpha-1}=\frac{\mathcal{B}(\alpha)}{\Sigma_{\rm crit}}\frac{ 2}{3-\alpha}\rho_c r_c^\alpha,
\end{equation}
we can then obtain the dimensionless convergence,
\begin{eqnarray}\label{eq:kappa}
&&\kappa(R)=\Sigma(R)/\Sigma_{\rm crit}= \nonumber\\
&&\left\{
           \begin{array}{ll}
             \displaystyle\frac{3-\alpha}{2}\left(\frac{b}{R}\right)^{\alpha-1}-\frac{3-\alpha}{ \mathcal{B}(\alpha)}\left(\frac{b}{r_c}\right)^{\alpha-1}\times  & \\\\
             \displaystyle
             \tilde{z}\left[F\left(\frac{\alpha}{2},1;\frac{3}{2};\tilde{z}^2\right)-F\left(\frac{\alpha_c}{2},1;\frac{3}{2};\tilde{z}^2\right)\right] &
\hbox{if $R\le r_c$} \\\\
             \displaystyle\frac{3-\alpha}{2}\left(\frac{b}{R}\right)^{\alpha-1}
& \hbox{if $R\ge r_c$}
           \end{array}
         \right. \nonumber\\
\end{eqnarray}
and the mean convergence within the radius $R$,
\begin{eqnarray}\label{eq:meankap}
&&\bar{\kappa}(R)=\frac{1}{\pi R^2}\frac{M_{2D}(R)}{\Sigma_{\rm crit}}= \nonumber\\
&&\left\{
           \begin{array}{ll}
              \displaystyle \left(\frac{b}{R}\right)^{\alpha-1}+\bar{\kappa}_0+\frac{2}{3}\frac{3-\alpha}{ \mathcal{B}(\alpha)}\left(\frac{b}{r_c}\right)^{\alpha-1}\left(\frac{r_c}{R}\right)^2\times & \\\\
             \displaystyle
             \tilde{z}^3\left[F\left(\frac{\alpha}{2},1;\frac{5}{2};\tilde{z}^2\right)-F\left(\frac{\alpha_c}{2},1;\frac{5}{2};\tilde{z}^2\right)\right] &
\hbox{if $R\le r_c$} \\\\
             \displaystyle \left(\frac{b}{R}\right)^{\alpha-1}+\bar{\kappa}_0
& \hbox{if $R\ge r_c$},
           \end{array}
         \right. \nonumber\\
\end{eqnarray}
where
\begin{equation}
\bar{\kappa}_0=\frac{1}{\pi R^2}\frac{m_0}{\Sigma_{\rm crit}}=-\frac{2}{ \mathcal{B}(\alpha)}\frac{\alpha-\alpha_c}{3-\alpha_c}\left(\frac{b}{r_c}\right)^{\alpha-1}\left(\frac{r_c}{R}\right)^2.
\end{equation}

From the above functions, we can see that the mass density profile considered here is actually a combination of a power-law mass distribution with a negative or positive mass distribution in the central region. Thus, the convergence can also be written as
\begin{equation}\label{eq:kappa12}
\kappa(R)=\Sigma(R)/\Sigma_{\rm crit}=\kappa_1(R)+\kappa_2(R),
\end{equation}
where
\begin{equation}\label{eq:kappa1}
\kappa_1(R)=\frac{3-\alpha}{2}\left(\frac{b}{R}\right)^{\alpha-1}
\end{equation}
is the power-law part, and $\kappa_2(R)$ denotes the second part in the central region
\begin{eqnarray}\label{eq:kappa2}
&&\kappa_2(R)=-\frac{3-\alpha}{ \mathcal{B}(\alpha)}\left(\frac{b}{r_c}\right)^{\alpha-1}\times \nonumber\\
&&\left\{
           \begin{array}{ll}
             \displaystyle \tilde{z}\left[F\left(\frac{\alpha}{2},1;\frac{3}{2};\tilde{z}^2\right)-F\left(\frac{\alpha_c}{2},1;\frac{3}{2};\tilde{z}^2\right)\right] &
\hbox{if $R\le r_c$} \\\\
             \displaystyle 0
& \hbox{if $R\ge r_c$}.
           \end{array}
         \right. \nonumber\\
\end{eqnarray}

\begin{figure}[htb!]
  \centering
  \includegraphics[width=0.45\textwidth]{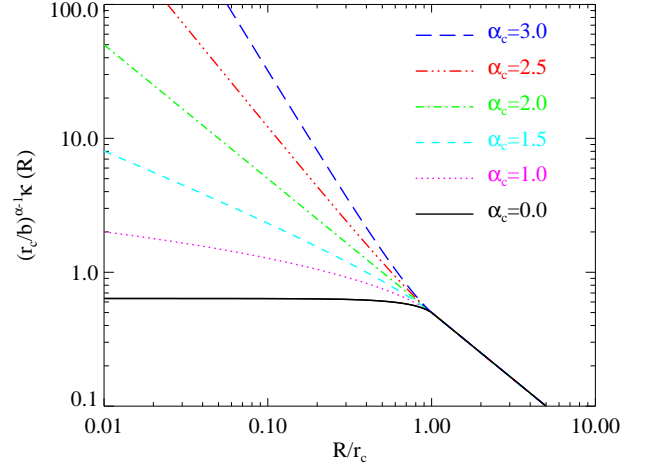}\\
  \caption{Convergence profiles of the BPL model with outer slope $\alpha=2$ and different inner slopes. The corresponding inner slope values are presented with the same colors as the lines.}\label{fig:kappa_calp}
\end{figure}

In Figure \ref{fig:kappa_calp}, we present several examples of the BPL convergence profiles with $\alpha=2$ and different values of $\alpha_c$. As shown, the inner part of the density profile can take the shape of either a flat core or a steep cusp, depending on the value of $\alpha_c$.

More generally, in order to describe the elliptical surface mass distributions, we can generalize the circular radius $R$ to an elliptical radius $R_{\rm el}=\sqrt{qx^2+y^2/q}$, where $q$ is the axial ratio of the isodensity ellipses. This definition of elliptical radius conserves the area and the total mass within $R_{\rm el}$.

\subsection{The Deflection Angle and Magnification}
The geometry of the deflection of a light ray can be described by the concise lens equation \citep{SWM2006book,2015A&A...580A..79T},
\begin{equation}\label{eq:lenseq}
  z_s=z-\alpha(z),
\end{equation}
which shows the mapping of a light ray from the position $z=x+{\rm i}y$ on the image plane to the position $z_s=x_s+{\rm i}y_s$ on the source plane, where $\alpha(z)=\alpha_x+{\rm i}\alpha_y$ is the scaled deflection angle and i denotes the imaginary unit. For an elliptical surface mass distribution, the corresponding complex conjugate of $\alpha(z)$ can be calculated using the formulation of BK75,
\begin{eqnarray}\label{eq:alp_star}
\alpha^*(z)&=&\frac{2}{z}\int_0^{R_{\rm el}}\frac{\kappa(R)R \ \dd R}{\sqrt{1-\zeta^2 R^2}}
\end{eqnarray}
with $\zeta^2=(1/q-q)/z^2$. In the special case of $q=1$, we find
\begin{equation}\label{eq:alp_sph}
\alpha^*(z)=\frac{R^2\bar{\kappa}(R)}{z}=\bar{\kappa}(R)z^*
\end{equation}
where $z^*$ is the complex conjugate of $z$.

Inserting Equation (\ref{eq:kappa12}) into Equation (\ref{eq:alp_star}), we then obtain the deflection field for the BPL model, \ie~
\begin{equation}
\alpha^*(z)=\alpha_1^*(z)+\alpha_2^*(z)
\end{equation}
with
\begin{equation}
\alpha_1^*(z)=\frac{R_{\rm el}^2}{z}\left(\frac{b}{R_{\rm el}}\right)^{\alpha-1}F\left(\frac{1}{2},\frac{3-\alpha}{2};\frac{5-\alpha}{2};\zeta^2 R_{\rm el}^2\right)
\end{equation}
for the SPL mass distribution $\kappa_1$ and
\begin{align}
\alpha_2^*(z)=&\frac{r_c^2}{z}\frac{3-\alpha}{\mathcal{B}(\alpha)}\left(\frac{b}{r_c}\right)^{\alpha-1}
\left[\frac{2}{3-\alpha_c}\mathscr{F}\left(\frac{3-\alpha_c}{2},\mathcal{C}\right)-\right. \nonumber\\
{}&\left.\frac{2}{3-\alpha}\mathscr{F}\left(\frac{3-\alpha}{2},\mathcal{C}\right)-
  \mathcal{S}_0\right]
\end{align}
for the complementary part $\kappa_2$, where
\begin{eqnarray}\label{eq:hyp3f2s}
&&\mathscr{F}(a,z)={_3F_2}\left(a,\frac{1}{2},1;a+1,\frac{3}{2};z\right)= \nonumber\\
&&\left\{
           \begin{array}{ll}
             \displaystyle \frac{1}{1-2a}\left[F(a,1;a+1;z)-2aF\left(\frac{1}{2},1;\frac{3}{2};z\right)\right] & \hbox{if $a \neq \frac{1}{2}$} \\\\
             \displaystyle \frac{{\rm Li}_2(\sqrt{z})-{\rm Li}_2(-\sqrt{z})}{2\sqrt{z}} & \hbox{if $a = \frac{1}{2}$}
           \end{array}
         \right. \nonumber\\
\end{eqnarray}
is a special case of the generalized hypergeometric function $_3F_2()$, where it is reduced to the Gauss hypergeometric function $F()$ here, and ${\rm Li}_2()$ denotes the Spence's function, \ie~${\rm Li}_2(z)=-\int_0^z\frac{\ln(1-t) dt}{t}$, and
\begin{eqnarray}\label{eq:series0}
&&\mathcal{S}_0(\alpha,\alpha_c,\tilde{z}_{\rm el},\mathcal{C})= \nonumber\\
&&\left\{
           \begin{array}{ll}
             \displaystyle \frac{1}{\sqrt{1-\mathcal{C}}}\sum_{n=0}^{\infty}\frac{(\frac{\alpha_c}{2})^{(n)}-(\frac{\alpha}{2})^{(n)}}{(\frac{3}{2})^{(n)}}\frac{2\tilde{z}_{\rm el}^{2n+3}}{2n+3} \times& \\\\
\displaystyle F\left(\frac{1}{2},\frac{2n+3}{2},\frac{2n+5}{2},\frac{\mathcal{C}\tilde{z}_{\rm el}^2}{\mathcal{C}-1}\right) &
\hbox{if $R_{\rm el}\le r_c$} \\\\
             \displaystyle 0
& \hbox{if $R_{\rm el}\ge r_c$}
           \end{array}
         \right. \nonumber\\
\end{eqnarray}
where $\mathcal{C}=r_c^2\zeta^2$, $\tilde{z}_{\rm el}=\sqrt{1-R_{\rm el}^2/r_c^2}$, and $x^{(n)}$ denotes the rising factorial of $x$.

The series $\mathcal{S}_0$ converges rapidly for pixels close to the break radius, due to the fact that $\tilde{z}_{\rm el}\to0$ as $R_{\rm el}\to r_c$, but the convergence becomes slower in the very central region where $\tilde{z}_{\rm el}\to1$. This is, however, not an issue, because there are only a few pixels in the central region of a lens. On the other hand, for real observations, the very central region usually suffers from larger noise than other regions because of the light contamination of the foreground lens. Therefore, in actual data analysis, the very central region of a lens could be masked to speed up the evaluation of deflection angles if there is no clearly visible central lensed image.

Given the analytical deflection field, it is straightforward to derive the lensing shear,
\begin{equation}\label{eq:gamma}
\gamma^*(z)=\frac{\partial \alpha^*}{\partial z}
\end{equation}
where
\begin{equation*}
\frac{\partial }{\partial z}=\frac{1}{2}\left[\frac{\partial }{\partial x}-{\rm i}\frac{\partial }{\partial y}\right]
\end{equation*}
is the Wirtinger derivative \citep{1926MathAnn97,1994A&A...284..285K}.

\begin{figure*}[htb!]
  \centering
  \includegraphics[width=0.9\textwidth]{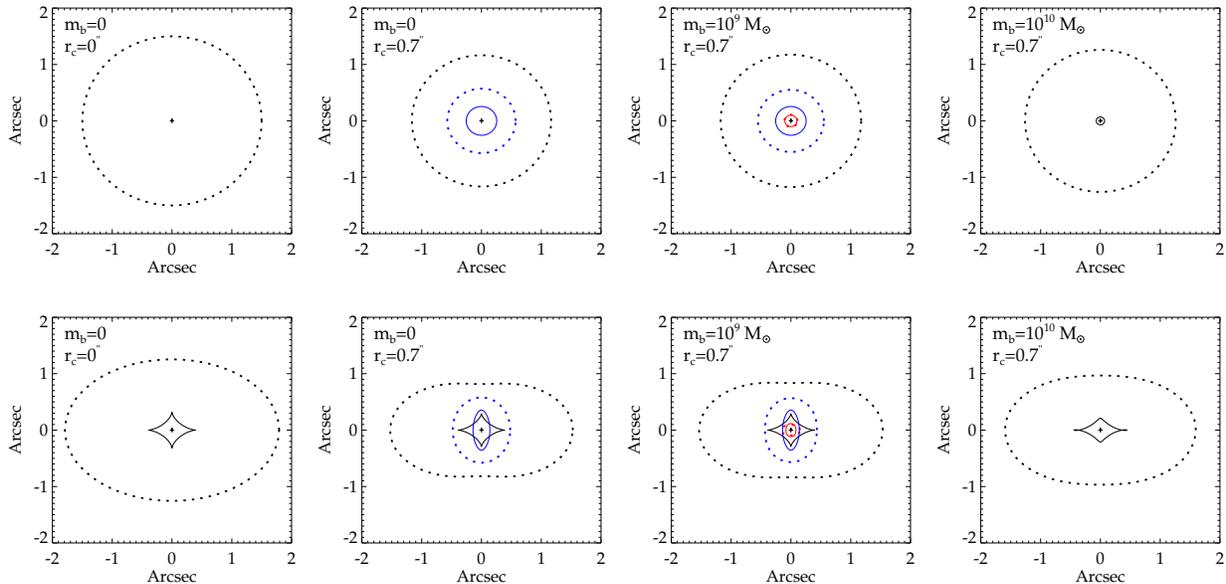}\\
  \caption{The critical curves (dotted) and caustics (solid with the same color as its critical curve) for the BPL density profiles with the same $b=1\farcs5$, $\alpha=2$, and $\alpha_c=0.6$ but with different ellipticity $q$, break radius $r_c$ or black hole mass $m_b$. The top panels are for spherical cases ($q=1$) and the bottom panels for elliptical cases ($q=0.7$). From left to right, the considered values of $m_b$ and $r_c$ are shown in the top-left corner of each panels. The small ``plus'' symbol in each panel marks the center of the lenses. Note that all the lenses are assumed to be at redshift $z_d\simeq0.178$ with a source at $z_s=0.6$.}\label{fig:cri_cau}
\end{figure*}

For the $\kappa_1$ part of the BPL model,
as shown in the paper of \citet{2015A&A...580A..79T}, the conjugate of the shear is
\begin{equation}
\gamma_1^*(z)=\frac{\partial \alpha_1^*}{\partial z}=(2-\alpha)\frac{\alpha_1^*}{z}-\kappa_1(z)\frac{z^*}{z}
\end{equation}
Similarly, the shear for the second part $\kappa_2$ is
\begin{equation}
\begin{split}
{\gamma_2^*(z)}&=\frac{\partial \alpha_2^*}{\partial z} \\
{}&=-\kappa_2(z)\frac{q|z|^2-(1+q^2)r_c^2}{qz^2-(1-q^2)r_c^2}+
2\frac{r_c^2}{z^2}\frac{3-\alpha}{\mathcal{B}(\alpha)}\left(\frac{b}{r_c}\right)^{\alpha-1}\times \\
{}&\left[\frac{2-\alpha_c}{3-\alpha_c}\mathscr{F}\left(\frac{3-\alpha_c}{2},\mathcal{C}\right)-
\frac{2-\alpha}{3-\alpha}\mathscr{F}\left(\frac{3-\alpha}{2},\mathcal{C}\right)-\mathcal{S}_2\right] 
\end{split}
\end{equation}
where
\begin{eqnarray}\label{eq:series2}
&&\mathcal{S}_2(\alpha,\alpha_c,\tilde{z}_{\rm el},\mathcal{C})= \nonumber\\
&&\left\{
           \begin{array}{ll}
             \displaystyle \frac{1}{(1-\mathcal{C})^{\frac{3}{2}}}\sum_{n=0}^{\infty}\frac{(\frac{\alpha_c}{2})^{(n)}-(\frac{\alpha}{2})^{(n)}}{(\frac{3}{2})^{(n)}}\frac{2n+2}{2n+3} \times& \\\\
\displaystyle \tilde{z}_{\rm el}^{2n+3}F\left(\frac{1}{2},\frac{2n+3}{2},\frac{2n+5}{2},\frac{\mathcal{C}\tilde{z}_{\rm el}^2}{\mathcal{C}-1}\right) &
\hbox{if $R_{\rm el}\le r_c$} \\\\
             \displaystyle 0
& \hbox{if $R_{\rm el}\ge r_c$}
           \end{array}
         \right. \nonumber\\
\end{eqnarray}
Note that the series $\mathcal{S}_2$ here also suffers from the same convergence problem as the series $\mathcal{S}_0$ in the very central region, where the calculations should be done with caution.

If there exists a central black hole, its effect on the shear field can be added. Based on Equations (\ref{eq:alp_sph}) and (\ref{eq:gamma}), the shear induced by a central black hole is
\begin{equation}
\gamma_{b}^*=-\frac{1}{\pi}\frac{m_b}{\Sigma_{\rm crit}}\frac{1}{z^2}
\end{equation}
where $m_b$ is the mass of the central black hole.

The magnification $\mu$ for the BPL density profile with a central black hole can be calculated according to
\begin{equation}\label{eq:lensmag}
  \mu^{-1}=(1-\kappa_1-\kappa_2)^2-|\gamma_1^*+\gamma_2^*+\gamma_{b}^*|^2 .
\end{equation}
By solving equation $\mu^{-1}=0$, the critical curves can then be found. The corresponding caustics can be obtained by mapping the critical curves on the lens plane to the source plane using the lens equation.

In Figure \ref{fig:cri_cau}, we present examples of the critical curves and caustics for several cases of BPL density profiles with $b=1\farcs5$, $\alpha=2$, and $\alpha_c=0.6$ but different ellipticity $q$, break radius $r_c$ or black hole mass $m_b$. The top and bottom panels are for the spherical ($q=1$) and elliptical cases ($q=0.7$), respectively. From left to right, the values of the core radius $r_c$ and black hole mass $m_b$ are shown for each case in the top-left corner. In order to demonstrate the effect of a black hole in units of $\msun$, all of the lenses are assumed to be at redshift $z_d\simeq0.178$ with a source at $z_s=0.6$.

For the spherical cases, it is shown in the first panel that there is only one critical curve for the singular isothermal sphere density profile, \ie~ the spherical power-law profile with $\alpha=2$. The second panel demonstrates that the radial critical curve will appear if there is a fairly flat core. Furthermore, the inclusion of a central massive black hole will produce the third critical curve in the region much closer to the center. However, if the central black hole is too massive, all of the inner critical curves will disappear and only the outmost tangential critical curve remains \citep[see examples in][]{2001MNRAS.323..301M}.

For the elliptical cases, the same number of critical curves are presented as the corresponding spherical cases. However, the critical curves become elliptical-like, and the caustics change accordingly. What is obvious is that the caustics for the outmost tangential critical curves are turned into astroid-like curves from a point or a circle.

The critical curves and caustics provide references to speculate on the image configurations. However, we know that the pattern of lensed images is not only sensitive to the lensing mass distributions but also the source properties. In Figures \ref{fig:arcs0}$-$\ref{fig:arcs2} in Appendix \ref{appendix:lenses}, we illustrate the complex dependence of the lensed images on the lens and source properties.

\subsection{The Velocity Dispersions}
In addition to the lensing observations, stellar kinematics can provide complementary information to the mass distributions of galaxies, especially the 2D kinematics from the integral-field spectroscopy \citep[IFS;][]{2015ApJ...798....7B,2016ARA&A..54..597C}. However, detailed IFS observations are currently only available for galaxies in the local universe whereas most of the SL systems are at redshift higher than 0.1 \citep{2008ApJ...682..964B,2015ApJ...803...71S}. For higher redshift galaxies, the most efficient way now is to measure the 1D velocity dispersions using the optical single-fiber spectroscopy. In this subsection, we investigate the velocity dispersions in detail based on the BPL model for the lensing mass.

We assume that the galaxies are spherical in the modeling of the dynamics of galaxies. According to the spherical Jeans equation and assuming a constant velocity anisotropy parameter $\beta$, the radial velocity dispersion for the stars is \citep{2008gady.book.....B}
\begin{equation}\label{eq:rvd}
\sigma_r^2(r)=\frac{G\int_r^\infty \dd\rpr j(\rpr)M(\rpr)(\rpr)^{2\beta-2}}{r^{2\beta}j(r)}
\end{equation}
where $j(r)$ is the 3D luminosity density profile related to the stellar number density profile and $M(r)$ corresponds to the total mass within the 3D radius $r$. The LOSVD at the projected radius $R$ is then given by
\begin{equation}\label{eq:losvd}
\sigma_\parallel^2(R)=\frac{\int_{-\infty}^\infty \dd Z~ j(r) (1-\beta R^2/r^2)\sigma_r^2(r)}{\int_{-\infty}^\infty \dd Z~ j(r)} .
\end{equation}

For single-fiber spectroscopic observations based on ground-based telescopes, the effect of fiber aperture and atmospheric seeing should be taken into account. In this case, the observed velocity dispersion can be modeled as
\begin{equation}\label{eq:al_losvd}
\langle\sigma_\parallel^2\rangle=\frac{\int_0^\infty \dd R~ R w(R) I(R) \sigma_\parallel^2(R)}{\int_0^\infty \dd R~ R w(R) I(R)}
\end{equation}
where $w(R)$ is the weighting function accounting for the fiber aperture size and the seeing effect, and
\begin{equation}\label{eq:Isurface}
  I(R)=\int_{-\infty}^\infty \dd Z~ j(r)
\end{equation}
is the surface brightness distribution. Rewriting Equation (\ref{eq:al_losvd}) by inserting Equation (\ref{eq:losvd}) and (\ref{eq:Isurface}), one gets
\begin{equation}\label{eq:al_losvd1}
\langle\sigma_\parallel^2\rangle=\frac{\int_0^\infty \dd R~ R w(R)\int_{-\infty}^\infty \dd Z~ j(r)(1-\beta R^2/r^2)\sigma_r^2(r) }{\int_0^\infty \dd R~ R w(R)\int_{-\infty}^\infty \dd Z~  j(r)}
\end{equation}
which is hereafter named as the aperture and luminosity (AL)-weighted LOSVD.

In practice, as discussed in \citet{2010ApJ...708..750S}, we can use a Gaussian smoothing function to approximate the weighting function $w(R)$ to some extent by assuming
\begin{equation}\label{eq:gauwr}
w(R)\approx\exp\left(-\frac{R^2}{2\sigma_{\rm fib}^2}\right)
\end{equation}
with
\begin{equation}\label{eq:sigfb}
\sigma_{\rm fib}\approx\sigma_{\rm see}\sqrt{1+\chi^2/4+\chi^4/40}
\end{equation}
and $\chi=R_{\rm fib}/\sigma_{\rm see}$,
where $\sigma_{\rm see}$ is the Gaussian standard deviation of the seeing function equivalent to the FWHM of the seeing divided by $2\sqrt{2\ln2}$.

If we adopt a Gaussian form of $w(R)$, \ie~ Equation (\ref{eq:gauwr}), by changing the order of integrations, Equation (\ref{eq:al_losvd1}) can be reduced to
\begin{eqnarray}\label{eq:allosvd}
&&\langle\sigma_\parallel^2\rangle= \nonumber \\
&&\frac{\int_0^\infty \dd r~ r^2 j(r)\sigma_r^2(r)
\left[\Phi\left(1,\frac{3}{2};-\frac{r^2}{2\sigma_{\rm fib}^2}\right)-\frac{2\beta}{3}\Phi\left(2,\frac{5}{2};-\frac{r^2}{2\sigma_{\rm fib}^2}\right) \right] }{\int_0^\infty \dd r~ r^2 j(r) \Phi\left(1,\frac{3}{2};-\frac{r^2}{2\sigma_{\rm fib}^2}\right)} \nonumber \\
\end{eqnarray}
with only 1D integrals, where the function $\Phi(a_1,a_2;-x)={\rm e}^{-x}\Phi(a_2-a_1,a_2;x)$ being the Kummer's confluent hypergeometric function.

Looking into Equation (\ref{eq:allosvd}), we notice that both $j(r)$ and $\sigma_{\rm fib}$ can be treated as known quantities: the $j(r)$ can be inferred from fitting to the surface brightness distribution, and $\sigma_{\rm fib}$ can be estimated from Equation (\ref{eq:sigfb}). The AL-weighted LOSVD $\langle\sigma_\parallel^2\rangle$ is a direct observable. The mass distribution that we are paying attention to is implicit in the radial velocity dispersion $\sigma_r^2(r)$.

To stay consistent with the BPL lensing mass model, we adopt the S\'ersic profile with a power-law inner density profile to describe the luminosity density profile of galaxies. It is the power-law S\'ersic (PL-S\'ersic) profile developed by \citet{2005MNRAS.362..197T}.  The 3D form of the PL-S\'ersic profile is written as
\begin{equation}\label{eq:jplsersic}
j(r)=\left\{
           \begin{array}{ll}
             \displaystyle j_c\left(r/r_c\right)^{-\alpha_c} &
\hbox{if $r\le r_c$} \\\\
             \displaystyle j_0\left(\frac{r}{s}\right)^{-u}\exp\left[-\left(\frac{r}{s}\right)^\nu\right]
& \hbox{if $r\ge r_c$}
           \end{array}
         \right.
\end{equation}
where
\begin{equation}
j_0=j_c\left(\frac{r_c}{s}\right)^{u}\exp\left[\left(\frac{r_c}{s}\right)^\nu\right] ,
\end{equation}
$j_c$ is the luminosity density at $r_c$, $s=R_{\rm eff}/k^n$ is a scale radius defined by the 2D effective radius $R_{\rm eff}$ for the single S\'ersic profile and the S\'ersic index $n$ (note that $k$ here is a function of $n$ and its expression can be found in \citealt{1999A&A...352..447C} and \citealt{2003ApJ...582..689M}), $\nu =1/n$, and $u=1-0.6097\nu+0.054635\nu^2$ \citep{1999MNRAS.309..481L,2001A&A...379..767M}.

The surface luminosity density profile corresponding to the 3D PL-S\'ersic profile is thus
\begin{eqnarray}\label{eq:iplsersic}
&&I(R)=2\int_{R}^{\infty}\frac{j(r) r \ \dd r}{\sqrt{r^2-R^2}}= \nonumber\\
&&\left\{
           \begin{array}{ll}

             \displaystyle2j_cr_c\tilde{z}F\left(\frac{\alpha_c}{2},1;\frac{3}{2};\tilde{z}^2\right)+2\int_{r_c}^{\infty}\frac{j(r) rdr}{\sqrt{r^2-R^2}}&
\hbox{~if $R\le r_c$} \\\\
             \displaystyle I_0\exp\left[-\left(\frac{R}{s}\right)^\nu\right]
& \hbox{~if $R\ge r_c$}
           \end{array}
         \right. \nonumber\\
&&
\end{eqnarray}
where
\begin{equation}\label{}
  I_0=2s j_0 \frac{\Gamma(\frac{3-u}{\nu})}{\Gamma(\frac{2}{\nu})} .
\end{equation}

In the following analyses, we simply assume that the BPL mass profile and the PL-S\'ersic light profile have the same break radius $r_c$ and the inner density slope $\alpha_c$. This assumption simplifies the velocity dispersion calculations.

Inserting Equation (\ref{eq:mass3d}) and (\ref{eq:jplsersic}) into Equation (\ref{eq:rvd}), we then derive an analytical form of the radial velocity dispersion for $r\ge r_c$, \ie~
\begin{eqnarray}\label{eq:radialvd}
\sigma_r^2(r)&=&G\frac{r^{-2\beta}}{J(r)}
\left\{\frac{4\pi\rho_cr_c^\alpha}{3-\alpha}\frac{s^\eta}{\nu}\Gamma\left[\frac{\eta}{\nu},\left(\frac{r}{s}\right)^\nu\right]+ \right.\nonumber\\
&&\left. m_0\frac{s^\lambda}{\nu}\Gamma\left[\frac{\lambda}{\nu},\left(\frac{r}{s}\right)^\nu\right]\right\} \nonumber\\
&=&\mathcal{A}\frac{r^{-2\beta}}{J(r)}
\left\{\frac{s^\eta}{\nu}\Gamma\left[\frac{\eta}{\nu},\left(\frac{r}{s}\right)^\nu\right]- \right.\nonumber\\
&&\left.r_c^{3-\alpha}\frac{\alpha-\alpha_c}{3-\alpha_c}\frac{s^\lambda}{\nu}\Gamma\left[\frac{\lambda}{\nu},\left(\frac{r}{s}\right)^\nu\right]\right\}
\end{eqnarray}
where $J(r)=r^{-u}\exp\left[-\left(\frac{r}{s}\right)^\nu\right]$, $\eta=2-u-\alpha+2\beta$, $\lambda=-1-u+2\beta$, and
\begin{equation}\label{}
\mathcal{A}=\frac{4\pi G\rho_cr_c^\alpha}{3-\alpha}=\frac{c^2}{2}\frac{D_s}{D_d D_{ds}}\frac{b^{\alpha-1}}{\mathcal{B}(\alpha)}
\end{equation}
showing the relation between the mass density profile and the lensing-related quantities.
For $r<r_c$, we obtain
\begin{eqnarray}\label{eq:radialvd1}
\sigma_r^2(r)&=&\frac{4\pi G\rho_cr_c^{\alpha_c}}{3-\alpha_c}\frac{r_c^\mu-r^\mu}{\mu}r^{\alpha_c-2\beta}+
\sigma_r^2(r_c)\left(\frac{r}{r_c}\right)^{\alpha_c-2\beta} \nonumber\\
&=&\mathcal{A}
\frac{3-\alpha}{3-\alpha_c}\frac{r_c^\mu-r^\mu}{\mu}\frac{r^{\alpha_c-2\beta}}{r_c^{\alpha-\alpha_c}}+
\sigma_r^2(r_c)\left(\frac{r}{r_c}\right)^{\alpha_c-2\beta} \nonumber\\
\end{eqnarray}
where $\mu=2-2\alpha_c+2\beta$. Note that $\sigma_r^2(r)$ is still finite even if $\mu\to0$, due to the fact that $\lim\limits_{\mu\to 0}\frac{r_c^\mu-r^\mu}{\mu}=\ln\frac{r_c}{r}$.

When $r_c=0$, the mass distribution has only the SPL part, and the light distribution follows the pure S\'ersic profile. In this case, the radial velocity dispersion becomes
\begin{equation}\label{}
\sigma_r^2(r)=\mathcal{A}\frac{r^{-2\beta}}{J(r)}\frac{s^\eta}{\nu}\Gamma\left[\frac{\eta}{\nu},\left(\frac{r}{s}\right)^\nu\right] .
\end{equation}

Generally, a black hole can exist at the center of a massive galaxy. In this case, the black hole can be regarded as a point mass which may have detectable effects on the velocity dispersion. With the PL-S\'ersic light profile considered here, the contribution of a black hole to the radial velocity dispersion is given by
\begin{eqnarray}\label{eq:vdbh}
&&\sigma_{{\rm b},r}^2(r)= \nonumber\\
&&\left\{
           \begin{array}{ll}
             \displaystyle Gm_{\rm b}\frac{r_c^{\lambda^\prime}-r^{\lambda^\prime}}{\lambda^\prime}r^{\alpha_c-2\beta}+
             \sigma_{{\rm b},r}^2(r_c)\left(\frac{r}{r_c}\right)^{\alpha_c-2\beta} &
\hbox{if $r\le r_c$} \\\\
             \displaystyle Gm_{\rm b}\frac{r^{-2\beta}}{J(r)}\frac{s^\lambda}{\nu}\Gamma\left[\frac{\lambda}{\nu},\left(\frac{r}{s}\right)^\nu\right]
& \hbox{if $r\ge r_c$}
           \end{array}
         \right. \nonumber\\
\end{eqnarray}
where $m_b$ denotes the mass of the central black hole and $\lambda^\prime=-1-\alpha_c+2\beta$.

\section{The simulated galaxies}
In this section, we shall describe the construction of mock galaxies used to verify the feasibility of the BPL lensing mass model on galaxy scales.

\subsection{The Illustris Simulation}\label{subsection:illusim}
The Illustris project consists of a series of large-scale hydrodynamic simulations, which incorporates various kinds of baryonic physics including gas cooling; star formation and evolution; feedback from active galactic nuclei, supernovae and supermassive black holes; and so on \citep{2014MNRAS.445..175G,2014MNRAS.444.1518V,2015A&C....13...12N}. We adopt in this work the highest resolution run, named the Illustris-1 simulation, to generate our mock catalogs.

The Illustris-1 simulation follows the dynamics of $1820^3$ dark matter particles, with $1820^3$ hydrodynamical cells initially in a periodic box with $75\hmpc$ a side. The mass resolution for dark matter particles is $\sim6.26\times10^6\msun$, and the baryonic matter has an initial mass resolution of $\sim1.26\times10^6\msun$.
Different gravitational softening lengths are applied to different types of particles. For dark matter particles, the softening length is fixed to be a comoving value of 1 \hkpc. For stars and black holes, the softening length is limited to a maximum value of 0.5 \hkpc in physical scale. For the gas cells, an adaptive softening length is defined according to the fiducial cell size and a floor given by the collisionless baryonic particles.

Focusing on the galaxies with stellar mass larger than $10^{10}h^{-1}\msun$, we finally extract 5343 galaxies in Illustris-1 at redshift zero. In order to generate mock galaxies with redshift consistent with the current observed strong lenses,
we artificially put the galaxies at redshift $z_d\simeq0.178$, which is close to the median redshift of lenses identified by the Sloan Lens ACS (SLACS) Survey \citep{2008ApJ...682..964B,2015ApJ...803...71S}. For the lens redshift $z_d\simeq0.178$, $1^{\prime\prime}$ corresponds to $3~\kpc$ for the cosmology adopted by the Illustris project \footnote{The cosmology used for Illustris is $\Omega_m=0.2726$, $\Omega_\Lambda=0.7274$, $\Omega_b=0.0456$, $\sigma_8=0.809$, $n_s=0.963$ and $h=0.704$.}.

\subsection{Classification of The Galaxy Types}
We classify the resolved $5343$ galaxies into two types based on their S\'ersic indices and stellar dynamical properties. The adopted S\'ersic index here is denoted as $n_0$ which is derived from the 3D fitting of the PL-S\'ersic profile to the stellar mass distribution (see Section \ref{subsection:3dfit} for details).

For the dynamical properties, we refer to the fraction of kinetic energy invested in the ordered rotation \citep{2012MNRAS.423.1544S,2017MNRAS.468.3883P}, which is given by
\begin{equation}
k_{\rm rot}=\frac{K_{\rm rot}}{K}=\frac{1}{K}\sum \frac{1}{2}m_i\left(\frac{\vec{j_i}\cdot\hat{J}}{|\vec{r_i}\times\hat{J}|}\right)^2
\end{equation}
where $m_i$ is the mass of the $i$th stellar particle, $\vec{j_i}=\vec{r_i}\times\vec{v_i}$ represents the specific angular momentum, $\hat{J}$ denotes the direction of the total angular momentum, and $K=\sum\frac{1}{2}m_i|\vec{v_i}|^2$ is the total kinetic energy of the galaxy.
In the calculation of $k_{\rm rot}$, in order to avoid possible bias due to satellites at the outskirts of the galaxy, we only consider the stellar particles within the spherical radius $r_{90}$, which is the radius enclosing $90\%$ of the total stellar mass.

We define the galaxies with $k_{\rm rot}<0.5$ and $n_0>1.5$ as ``elliptical'' galaxies while the rest as ``disk'' galaxies. We thus obtain $1362$ elliptical galaxies which make up about $25\%$ of galaxies with stellar mass larger than $10^{10}h^{-1}\msun$. This fraction of elliptical galaxies is roughly consistent with that found in observations  \citep[see][]{2011MNRAS.412..246V,2012MNRAS.419.1324D,2013MNRAS.433.2986W}.

\subsection{The Surface Mass and Light Distribution}
The surface mass distribution for a galaxy is obtained by projecting its 3D mass distribution along the $x$-direction. All of the matter components are taken into account for a galaxy, including the dark matter, stars, gas, and black holes.

Among the lensing-related quantities, what we are more interested in is the convergence map. We thus generate convergence maps by scaling the surface mass distributions directly using the critical surface mass density $\Sigma_{\rm crit}$. For a lens system with the lens redshift $z_d=0.178$ and source redshift $z_s=0.6$, we have $\Sigma_{\rm crit}\simeq4.0\times10^{15}\msun/\mpc^{2}$ for the Illustris-adopted cosmology.

By looking into the 5343 convergence maps, we find that only a small fraction of galaxies can produce observable SL images. For example, there are only $358$ galaxies (most of them are elliptical galaxies, as also seen in observations) with Einstein radius larger than $0\farcs5$. The lensing probabilities for these Illustris galaxies are likely to be underestimated because of the smoothing effect of gravitational softening on the density profiles (see Section \ref{subsection:illusim} for the smoothing lengths for the stellar and dark matter particles). In the following analyses, we mainly focus on the density profile fittings to the galaxies regardless of their lensing probabilities.

As for the light distributions, they are directly approximated using the stellar mass distributions with a constant stellar mass-to-light ratio. For further simplification, we use the stellar mass distribution to represent the light distribution, as we are more concerned about the general shape of the light distribution rather than its total luminosity or amplitude. So, hereafter, when we refer to the ``light distribution'' in this paper, it is actually the mass distribution of the stellar matter. The ``mass distribution'' thus represents the overall mass distribution including all the matter components.

Consistent with the {\it Hubble} images, the resolution of $0\farcs05$ is taken to pixelize the mass and light distributions where the triangular-shaped cloud algorithm is applied \citep{1981CSUP}.

\subsection{The AL-weighted LOSVD}
We calculate the AL-weighted LOSVDs for the Illustris galaxies as follows:
\begin{equation}\label{eq:simvd}
\sigma_{\parallel,\rm sim}^2=\frac{\sum\omega_i v_{\parallel,i}^2}{\sum\omega_i}-\left(\frac{\sum\omega_i v_{\parallel,i}}{\sum\omega_i}\right)^2
\end{equation}
where $v_{\parallel,i}$ is the velocity component parallel to the line of sight for the $i$th stellar particle and $\omega_i=m_i \exp\left(-\frac{R_i^2}{2\sigma_{\rm fib}^2}\right)$ is the weighting function, with $m_i$ and $R_i$ denoting the mass and projected radius of the $i$th stellar particle, respectively. In this paper, we use $\sigma_{\rm fib}=1.15$, which is derived according to Equation (\ref{eq:sigfb}) in view of the fiber radius $1\farcs5$ and a typical seeing of $1\farcs69$ for the SLACS lenses.

In the theoretical modeling of the AL-weighted LOSVD, a constant velocity anisotropy parameter $\beta$ is assumed for each galaxy. In simulations, $\beta$ is evaluated by the global anisotropy \citep{2007MNRAS.379..418C},
\begin{equation}\label{eq:beta}
  \beta=1-\frac{\Pi_{\theta\theta}+\Pi_{\phi\phi}}{2\Pi_{rr}}
\end{equation}
with $\Pi_{kk}=\sum_{i=1}^{N} M_i\sigma_{k,i}^2$, the total energy from random motions along the $k$-direction (\ie~ the $r$, $\theta$ and $\phi$ directions in the spherical coordinate system), where the sum runs over all the radial bins within the radius $r_{90}$, and $M_i$ and $\sigma_{k,i}$ are the total mass and velocity dispersion along the $k$-direction in the $i$th radial bin, respectively.

Based on Equation (\ref{eq:beta}), we find that the global anisotropy $\beta$ is not sensitive to the radial binning, because of the mass weighting in each bin, although there exists radial variation for the $\beta$ in general. For instance, the estimated values of $\beta$ are almost the same for $30$ bins linearly spaced in $r$ starting from $r_{\rm min}=0 \ \kpc$, or logarithmically spaced starting from $r_{\rm min}= 0.15 \ \kpc$. We use the $\beta$ value estimated from the $30$ linear bins for each galaxy in the following analyses.

\section{Density Profile Estimations}
This section shows the application of the BPL model to the mass density profile estimations. Both the 3D and 2D fitting procedures are investigated.

\begin{deluxetable*}{l|l|l}
\tablecaption{3D and 2D density profile fittings \label{tab:table}}
\tablehead{
\colhead{Method} & \colhead{$\chi^2$ performance} & \colhead{Fitting range or area} }
\startdata
3D (Radial) & $\chi_{\bar\rho}^2+\chi_{\rho}^2+\chi_{j}^2$ using logarithmic radial bins & $0.3~\kpc~\rightarrow~r_{90}$ \\
\hline
2D (Elliptical) & $\chi_{\bar\kappa}^2+\chi_{\kappa}^2+\chi_{I}^2$ on pixelated maps & $14''\times14''$ or $6''\times6''$ (central 9 pixels are masked)\\
\hline
2D Radial & $\chi_{\bar\kappa}^2+\chi_{\kappa}^2+\chi_{I}^2$ using logarithmic radial bins & $0.3~\kpc~\rightarrow~R_{90}$ \\
\enddata
\tablecomments{For the 3D and 2D radial fittings, only the innermost bin is considered for the calculation of $\chi_{\bar\rho}^2$ and $\chi_{\bar\kappa}^2$. For the 2D elliptical fittings, only the 16 pixels around the central 9 pixels are used for the $\chi_{\bar\kappa}^2$.}
\end{deluxetable*}

\subsection{Fitting the 3D Density Profiles}\label{subsection:3dfit}
For the 3D fittings to the mass and light distributions, $30$ radial bins are equally spaced logarithmically in the range from $r=0.15 \ \kpc$ to $r_{90}$. The radius of the $i$th bin is denoted as $r_i$, which corresponds to the mean of $\lgg(r)$ in the $i$th bin. We then compute the spherically averaged mass density $\rho_i$ and light density $j_i$ at $r_i$ in the $i$th bin, and the mean mass density $\bar\rho_i(<r_i)$ within $r_i$. In view of the larger density uncertainties in the central region due to fewer particles and the possible center offsets between the different matter components, we only use the radial bins with radius larger than $0.3 \ \kpc$ for the 3D fittings to avoid systematics.

The total $\chi^2$ to be minimized is made of three pieces, \ie
\begin{equation}\label{eq:chi3d}
\chi_{\rm 3D}^2=\chi_{\bar\rho}^2+\chi_{\rho}^2+\chi_{j}^2
\end{equation}
where
\begin{eqnarray}
\chi_{\bar\rho}^2&=& \sum_{i=1}^{1}\left[\ln\bar\rho_{i}-\ln\bar\rho_{i,\rm BPL}(\rho_c,r_c,\alpha_c,\alpha, m_b)\right]^2 \nonumber\\
\chi_{\rho}^2&=&\sum_i\left[\ln\rho_i-\ln\rho_{i,\rm BPL}(\rho_c,r_c,\alpha_c,\alpha)\right]^2 \\
\chi_{j}^2&=&\sum_i \left[\ln j_i-\ln j_{i,\text{PL-S\'ersic}}(j_c,r_c,\alpha_c,R_{\rm eff},n)\right]^2 , \nonumber
\end{eqnarray}
with $\chi^2$ denoting the mean density profile $\bar\rho$ for only the innermost bin used to constrain the black hole mass, $\rho$ the mass density profile, and $j$ the light density profile.

Note that the mass and light density profile models share the same parameters $r_c$ and $\alpha_c$, because we assume that they have compatible inner density profiles. There are in total eight free parameters when the center is fixed, \eg~at the particle position with minimum gravitational potential energy. In order to estimate the S\'ersic index $n_0$ to better clarify the galaxy types, the light density profiles are also fitted by just minimizing the $\chi_{j}^2$.

\subsection{Fitting the Surface Density Profiles}
For 2D fittings, the maximum field of view (FoV) is chosen to be $14''\times14''$ with $281\times281$ pixels. This FoV is large enough for galaxy-scale SL observations because the Einstein radii are typically less than $3''$ for almost all the galaxy-scale lenses that have been discovered so far.

Similar to the 3D fittings, the 2D fittings are carried out by minimizing the following $\chi^2$,
\begin{eqnarray}\label{eq:chi2dm}
\chi_{\rm 2D}^2 &=& \chi_{\bar\kappa}^2+\chi_{\kappa}^2+\chi_{I}^2
\end{eqnarray}
with
\begin{eqnarray}
\chi_{\bar\kappa}^2 &=& \sum_i\left[\bar\kappa_i-\bar\kappa_{i,\rm BPL}(b,r_c,\alpha_c,\alpha, m_b)\right]^2 \nonumber\\
\chi_{\kappa}^2 &=& \sum_i\left[\kappa_i-\kappa_{i,\rm BPL}(b,r_c,\alpha_c,\alpha,q,\phi)\right]^2 \\
\chi_{I}^2 &=& \sum_i\omega_I^2\left[I_i-I_{i,\text{PL-S\'ersic}}(j_c,r_c,\alpha_c,R_{\rm eff},n,q_I,\phi_I)\right]^2 \nonumber
\end{eqnarray}
where the data value of $\bar\kappa_i(<R_{{\rm el},i})$ at the $i$th pixel is the mean of the pixels of convergence within the elliptical radius $R_{{\rm el},i}=\sqrt{q x_i^2+y_i^2/q}$, and $\kappa_i$ and $I_i$ are the convergence and surface brightness at the $i$th pixel, respectively. In $\chi_{I}^2$, $\omega_I=\frac{M}{L}\frac{1}{\Sigma_{\rm crit}}$ is applied to make the amplitude of the surface brightness comparable to the convergence map. As previously mentioned, the light intensity $I$ here is actually the surface mass distribution of stars, \ie~ with $M/L=1$. As for the model parameters, $b$ is the scale radius defined in Equation (\ref{eq:brhoc}), $q$ and $\phi$ are the axis ratio and position angle of the surface mass distribution, respectively, while $q_I$ and $\phi_I$ are for the surface brightness distribution. In addition to the center $(x_0,y_0)$, there are in total $14$ free parameters in the 2D elliptical fitting to a galaxy.

Note that for the $\chi_{\rm 2D}^2$ defined above, the central nine pixels of the convergence maps are masked in order to reduce the possible influence of the black hole mass on the central mass distribution due to pixelization. For the $\chi_{\bar\kappa}^2$ which can constrain the black hole mass, only the 16 pixels surrounding the central 9 pixels are considered in the 2D fittings.

As we know, the actual mass density profiles are steeper at larger radius and likely to be truncated at a certain radius \citep[\eg][]{1997ApJ...490..493N,1999MNRAS.307..162S,2017MNRAS.468.2345D}. The BPL model proposed in this paper is a model for describing the mass distribution in the relatively central region of galaxies. The BPL fittings to the surface mass distributions may be sensitive to the fitting area. To examine the effect of the fitting area, we also pay attention to an FoV of $6''\times6''$ in the 2D elliptical fittings.

In order to quantify the possible bias more accurately, we further inspect the ``2D Radial'' fittings to the azimuthally averaged surface density profiles, which are directly calculated based on the matter particles. For the 2D radial fittings, the $\chi_{\rm 2D}^2$ expression defined in Equation (\ref{eq:chi2dm}) is adopted but with
\begin{eqnarray}
\chi_{\bar\kappa}^2 &=& \sum_{i=1}^{1}\left[\ln\bar\kappa_{i}-\ln\bar\kappa_{i,\rm BPL}(b,r_c,\alpha_c,\alpha, m_b)\right]^2 \nonumber\\
\chi_{\kappa}^2 &=& \sum_i\left[\ln\kappa_i-\ln\kappa_{i,\rm BPL}(b,r_c,\alpha_c,\alpha)\right]^2 \\
\chi_{I}^2 &=& \sum_i\left[\ln I_i-\ln I_{i,\text{PL-S\'ersic}}(j_c,r_c,\alpha_c,R_{\rm eff},n)\right]^2 \nonumber
\end{eqnarray}
where $i$ indicates the $i$th radial bin and the center is fixed at the position corresponding to the 3D center. A total of $30$ bins are equally spaced in logarithmic scale in the range of $0.15~\kpc$ to the projected $90\%$ light radius $R_{90}$. Only the radial bins with a projected radius $R$ larger than $0.3~\kpc$ are used in the analysis. Similar to the 3D fittings, only the innermost bin is adopted to calculate $\chi_{\bar\kappa}^2$. For a reference, table \ref{tab:table} briefly summarizes all the fitting methods investigated above.

\section{Results}
In this section, we present the results of the density profile fittings to the simulated Illustris galaxies. More attention is paid to the BPL mass model fittings. We also investigate the AL-weighted LOSVDs, which are modeled by the BPL mass and PL-S\'ersic light density profiles.

\subsection{The 3D Fittings}\label{subsection:3dresult}
\begin{figure*}[htb!]
  \centering
  \includegraphics[width=0.9\textwidth]{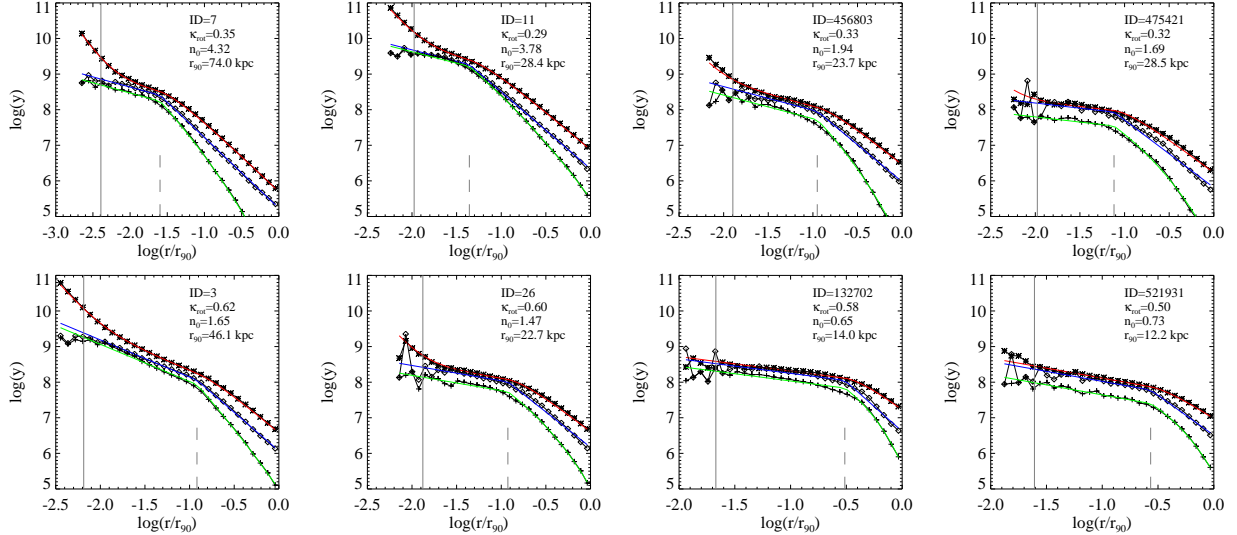}\\
  \caption{The 3D fittings to the mean mass density $\bar\rho$ (asterisks), mass density $\rho$ (diamonds), and light (or stellar mass) density $j$ (pluses) profiles in units of $\msun/\kpc^3$. The galaxies in the top and bottom panels are for the elliptical and disk galaxies, respectively. The red and blue lines show the BPL model fittings to the $\bar\rho$ and $\rho$, respectively. The green lines show the PL-S\'ersic fittings to the light density profiles. The vertical solid line in each panel presents the minimum fitting radius $0.3~\kpc$ and the vertical dashed line indicates the location of the break radius $r_c$. The subhalo ID, energy fraction invested in ordered rotation $\kappa_{\rm rot}$, the S\'ersic index $n_0$ used for galaxy-type classification (from the pure $\chi^2_j$ fitting rather than from the green line), and $90\%$ light radius $r_{90}$ are presented in the top-right corner of each panel.}\label{fig:3dfitex}
\end{figure*}

Figure \ref{fig:3dfitex} displays the 3D fittings for a couple of typical galaxies. We notice that, for most of the galaxies in Figure \ref{fig:3dfitex}, the BPL and PL-S\'ersic models work equally well to describe the relevant density profiles within $r_{90}$. However, obvious deviations exist for some galaxies in the extremely central region or around the break radius. The large fluctuations in the central region are likely due to the limited resolution of the Illustris simulation and the possible center offset between different matter components. The deviations around the break radius are expected because the BPL model is a piecewise function which is continuous but not smooth at the break radius.

\begin{figure*}[htb!]
  \centering
  \includegraphics[width=0.9\textwidth]{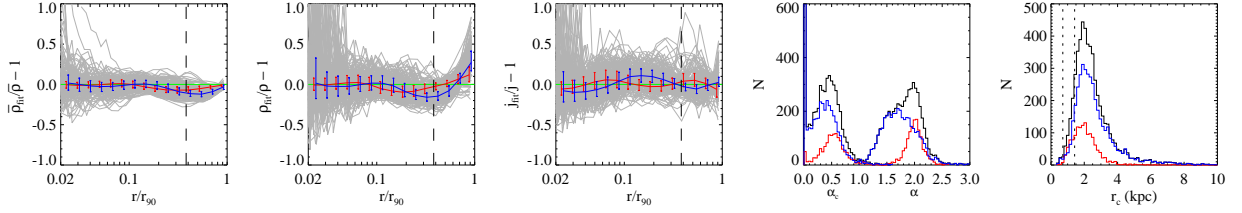}\\
  \caption{Statistical results of the 3D fittings. The first panel shows the relative deviations of $\bar{\rho}_{\rm fit}$ from $\bar{\rho}$ as a function of radius scaled by the $90\%$ light radius $r_{90}$. The second and third panels show the relative deviations of the fittings to the mass density ($\rho$) and light density ($j$) profiles, respectively. For clarity, results for 200 randomly selected galaxies are presented in the left three panels, where each gray line corresponds to one galaxy. The red and blue lines with error bars show the median deviations for all $1362$ elliptical galaxies and all $3981$ disk galaxies, respectively. The error bars indicate the range of the first and third percentiles in each bin. For reference, the median of the half-light radii for all galaxies is marked by the vertical dashed lines. The fourth panel displays the distributions of the inner slope $\alpha_c$ and outer slope $\alpha$. The rightmost panel displays the distributions of the break core radius $r_c$. The two vertical dotted lines indicate the softening lengths for the stars ($0.5\hkpc$) and dark matter ($1\hkpc$), respectively. In the last two panels, the black histograms are for all the galaxies while the red and blue histograms are for the elliptical and disk galaxies, respectively.}\label{fig:3djoint}
\end{figure*}

In Figure \ref{fig:3djoint}, we show the statistical results of the 3D fittings. As shown in the left panel of Figure \ref{fig:3djoint}, the biases of the BPL fittings to $\bar{\rho}(<r)$ (corresponding to the total mass distribution) are typically less than $5\%$ for elliptical galaxies and $10\%$ for disk galaxies. If we look into the corresponding fittings to the volume density profiles $\rho(<r)$ shown in the second panel, an obviously increasing trend is found for the biases at larger radius. The reason is that the true density profile is steeper at larger radius whereas the slope of the BPL model here is mainly determined by the mass distribution in the relatively inner region. Thus, the density profile at a larger radius tends to be overestimated by the BPL model. The biases can reach $30\%$ and $10\%$ at $r_{90}$ from a negative bias of about $-10\%$ for the disk and elliptical galaxies, respectively.

The third panel shows the relative deviations of the PL-S\'ersic profile fittings from the light distributions. It is shown that the deviation fluctuations are somewhat larger than the BPL fittings to the mass distributions, but still reasonable in consideration of the interplay between the mass and light density profiles in the $\chi_{\rm 3D}^2$ fittings.

The slope distributions in the fourth panel illustrate the obvious differences between the inner and outer density profiles within $r_{90}$. The inner slope is about $0.5$ for both the elliptical and disk galaxies. The outer slope is about $2$ with a scatter of $\sim0.16$ for the elliptical galaxies. However, for disk galaxies, the fitted outer slope is much flatter than for elliptical galaxies and has a larger dispersion. In the last panel, we find that the distributions of the break radii, if normalized, are nearly the same for elliptical and disk galaxies. The modes of the $r_c$ distributions are about $2~\kpc$, which is larger than the softening length of dark matter particles.

To sum up, the 3D fittings inspected in this subsection demonstrate that the BPL model is feasible to describe the mass density profiles of Illustris galaxies within a certain radius and the PL-S\'ersic model is also good enough to measure the light density profiles.

\subsection{2D Fittings}
\begin{figure*}[htb!]
  \centering
  \includegraphics[width=0.9\textwidth]{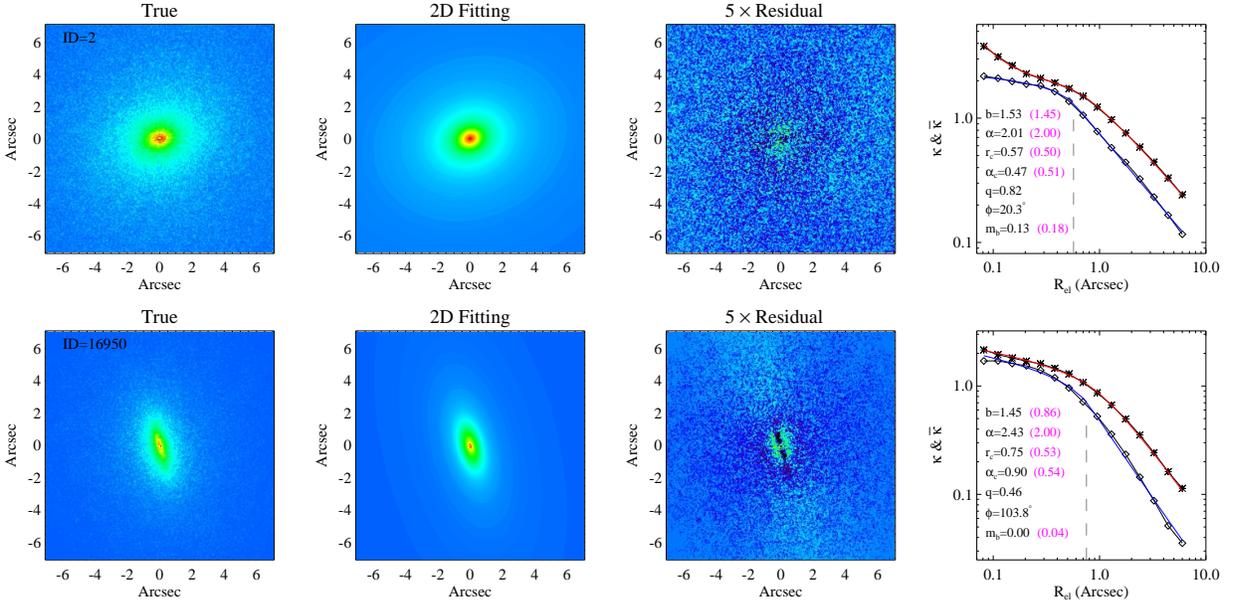}\\
  \caption{Two examples of 2D elliptical fittings to the convergence maps with the FoV of $14''\times14''$. The top and bottom rows are for an elliptical galaxy ($\rm{ID}=2$) and a disk galaxy ($\rm{ID}=16950$), respectively. The first and second columns show the true convergence maps and the BPL fits, respectively, while the third column shows the residuals (amplified $5$ times for illustration). The last column displays the convergence $\kappa$ and the averaged convergence $\bar{\kappa}(<R_{\rm el})$ as a function of the elliptical radius $R_{\rm el}$ which is adopted to be consistent with that adopted in the 2D elliptical fittings. The asterisks and diamonds denote the data points of $\bar{\kappa}(<R_{\rm el})$ and $\kappa$ respectively. The red and blue lines are the fitting profiles. The best-fit parameter values are shown by the black numbers in the panels where the black hole mass $m_b$ is in units of $10^{10}\msun$. The break radius for each galaxy is marked by the vertical dashed line. The parameter values for the 3D fittings are also shown by the magenta numbers for a reference.}\label{fig:2dfitex}
\end{figure*}

Figure \ref{fig:2dfitex} shows two examples of the 2D elliptical fittings to the convergence maps with an FoV of $14''\times14''$. As shown in Figure \ref{fig:2dfitex}, the BPL model can fit very well the 2D mass distributions of the two galaxies inspected here. For both galaxies, the residuals are sufficiently small, and from the rightmost panels, we can see that the BPL model can provide excellent fits to the elliptical radial density profiles.

By comparing the parameter values derived from 2D fittings (black numbers) and 3D fittings (magenta numbers), we find that they are not always consistent, especially for the disk galaxies. This can be attributed to the projection effect caused by the limitation of the BPL model and the nonspherical shape of galaxies. We present more detailed comparisons between the 2D and 3D BPL fittings in Section \ref{subsection:compar2d3d}.

\begin{figure*}[htb!]
  \centering
  \includegraphics[width=0.9\textwidth]{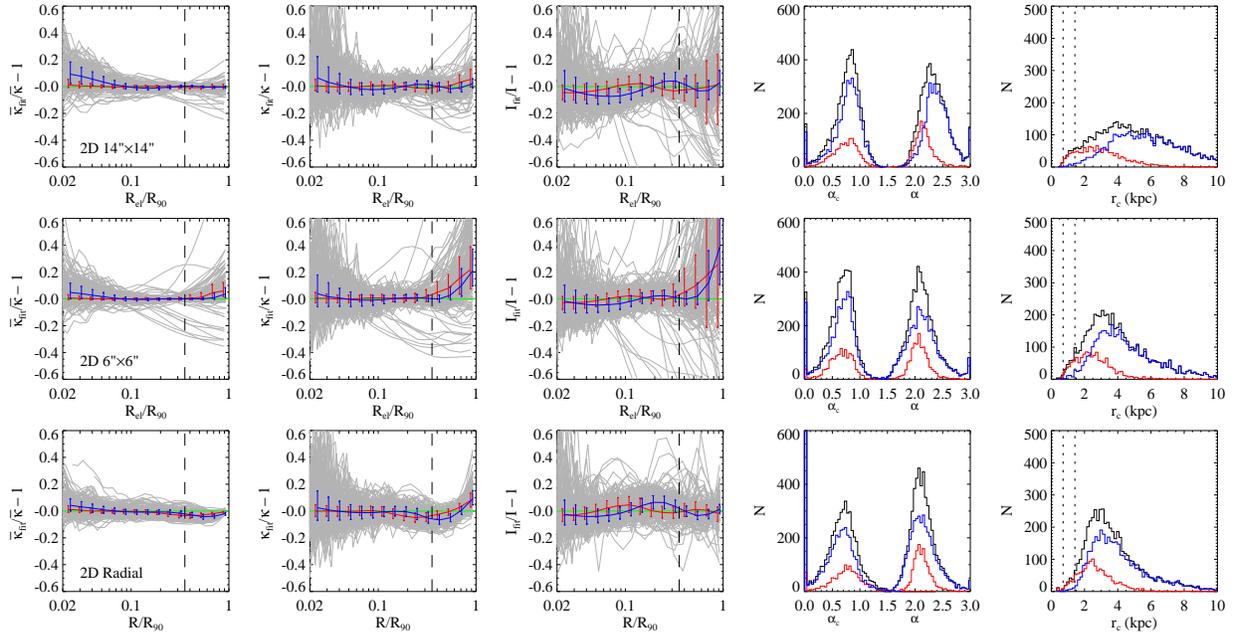}\\
  \caption{The statistical results of the 2D fittings to the mean convergence $\bar{\kappa}$, convergence $\kappa$ and surface brightness $I$. The top, second, and bottom panels are for the 2D fittings with FoV of $14''\times14''$ and $6''\times6''$ and the ``2D Radial'' fittings, respectively. The relative deviations $\bar{\kappa}_{fit}/\bar{\kappa}-1$, ${\kappa}_{fit}/{\kappa}-1$ and $I_{fit}/I-1$ for $200$ randomly selected galaxies are plotted with gray lines in the left three columns as a function of elliptical radius $R_{\rm el}$ (for the 2D elliptical fittings) or spherical radius $R$ (for the 2D radial fittings) scaled by the projected $90\%$ light radius $R_{90}$. The vertical dashed lines indicate the median of the projected half-light radii for all the galaxies. In the last two columns, the distributions of the parameter values of $\alpha_c$, $\alpha$, and $r_c$ are presented. The vertical dotted lines in the last column mark the softening lengths of the stars ($0.5\hkpc$) and dark matter ($1\hkpc$). The colored lines and histograms have similar meanings to those shown in Figure \ref{fig:3djoint}, but for 2D density profiles.}\label{fig:2dstat}
\end{figure*}

We now move to the statistical results of the 2D fittings, which are displayed in Figure \ref{fig:2dstat}. It is shown that, for all three methods of the 2D fitting, the BPL model can estimate the mean convergence $\bar{\kappa}$ maps of elliptical galaxies very well with a negligible bias within the radius $R_{90}$. However, for the disk galaxies, the mean convergence maps tend to be overestimated in the inner region, especially for the fittings with larger FoV.

For the BPL fittings to the convergence maps, as expected, biases always exist at much larger radius both for elliptical and disk galaxies. A smaller FoV may cause a worse fitting of convergence in the outer region. However, as indicated by the left panels, the averaged convergence within $R_{90}$ is not very susceptible to the overestimation of the convergence at a relatively larger radius.

As for the PL-S\'ersic fittings shown in the third column, the scatters are relatively larger for the 2D elliptical fittings. One reason is that the adopted FoV is not large enough for some massive galaxies. Therefore, the large deviations may emerge for massive galaxies when the radius is scaled by $R_{90}$.

In the fourth column of Figure \ref{fig:2dstat}, it is shown that the inner and outer density profiles can be still clearly separated by the slopes estimated from 2D fittings. The slope distributions are more consistent between the elliptical and disk galaxies. However, one may realize that both the inner and outer slopes are systematically higher than those from 3D fittings. For example, the modes of the inner slope distributions increase to about $0.8$ from $0.5$ for both the elliptical and disk galaxies. The mode of the outer slope distribution is biased to be about $2.3$ (for 2D $14''\times14''$ fittings ) or $2.1$ (for 2D $6''\times6''$ and radial fittings) for disk galaxies. However, for the elliptical galaxies, the outer slope distribution is only slightly biased.

In the last column, we display the break radius distributions, which are much wider than the distributions from 3D fittings, especially for the disk galaxies. The break radii for elliptical galaxies are systematically smaller than those for disk galaxies

Based on the results presented in this subsection, we realize that the performance of the 2D BPL fittings is sensitive to the binning or weighting methods. Even so, the inner and outer density profiles can be clearly differentiated whichever fitting method is used. More importantly, we find that the mean convergence maps can be recovered very well by the BPL fittings, especially for the elliptical galaxies.

\subsection{Comparisons between 2D and 3D Fittings}\label{subsection:compar2d3d}
As a model used to describe the relatively central region of galaxies, both the 2D and 3D BPL model fittings tend to overestimate the true mass density profiles outside a certain radius, because the outer slope of the BPL model is fixed while the true mass density profile usually decreases more and more rapidly with increasing radius. In addition to the nonspherical shape of galaxies, inconsistency must exist between the 2D and 3D BPL model fittings.

\begin{figure*}[htb!]
  \centering
  \includegraphics[width=0.9\textwidth]{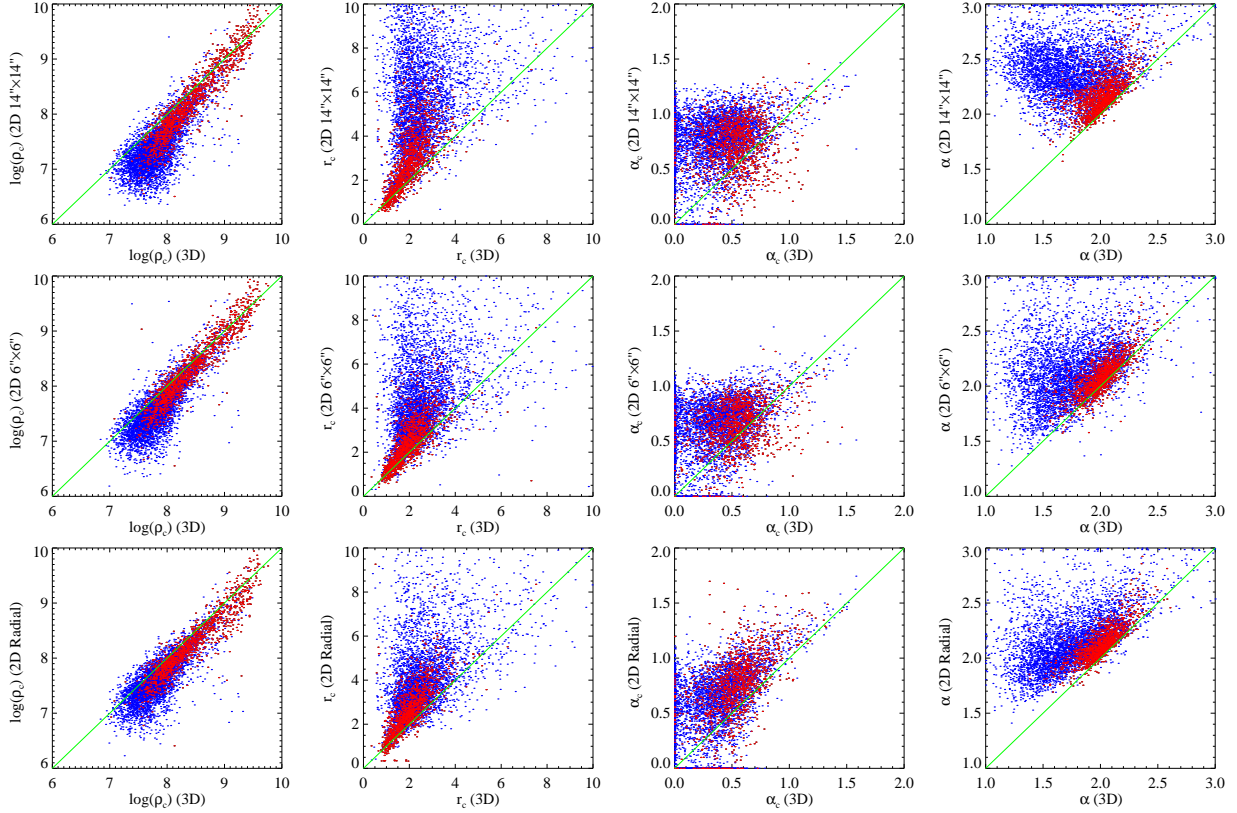}\\
  \caption{One-to-one comparisons of parameter values of $\rho_{c}$, $r_c$, $\alpha_c$ and $\alpha$ between 2D and 3D fittings, where $\rho_{c}$ is in units of $\msun/\kpc^3$ and $r_c$ is in kiloparsecs. The top and second rows show the comparisons of the 2D elliptical fittings with FoV of $14''\times14''$ and $6''\times6''$ to the 3D fittings, respectively. The bottom panels show the comparisons between the 2D radial and 3D fittings. In the plots, the red and blue dots are, respectively, for elliptical and disk galaxies. The green line in each panel indicates the identity line for a reference.}\label{fig:2d3d11}
\end{figure*}

Figure \ref{fig:2d3d11} presents the one-to-one comparisons of parameter values of $\rho_{c}$, $r_c$, $\alpha_c$, and $\alpha$ between the 2D and 3D fittings, where the $\rho_c$ for a 2D fitting is directly derived from the $b$, $\alpha$ and $r_c$ according to Equation \ref{eq:brhoc}. In these comparisons, the parameter values of 2D elliptical fittings are adopted directly to be compared by ignoring the nonspherical shape of galaxies.

From Figure \ref{fig:2d3d11}, we can realize that the 2D and 3D parameter values are not always strongly correlated, where large scatters and biases may exist, especially for the disk galaxies. We find, compared to the 3D fittings, that the values of $\rho_{c}$ estimated by 2D fittings are slightly underestimated in general but with a larger break radius $r_c$, and steeper inner and outer slopes.

\begin{figure*}[htb!]
  \centering
  \includegraphics[width=0.75\textwidth]{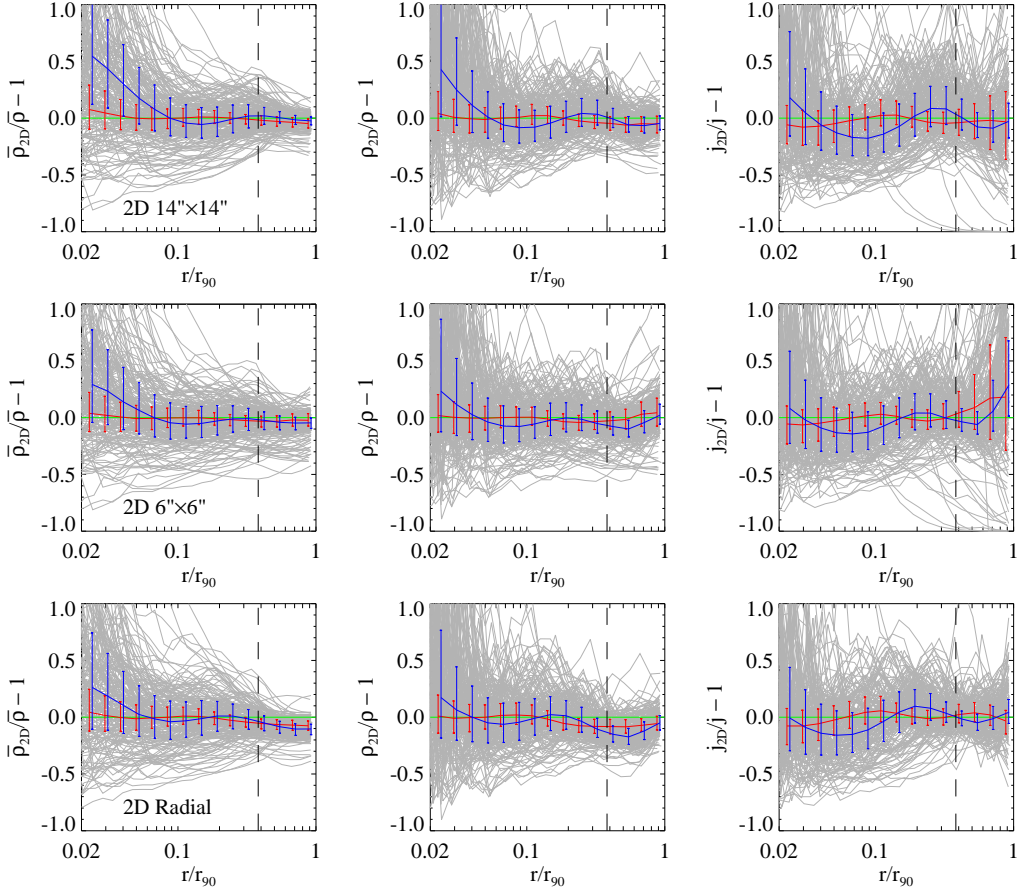}\\
  \caption{The relative differences between the 3D density profiles (\ie~$\bar{\rho}_{2D}$, $\rho_{2D}$, and $j_{2D}$) deprojected from the 2D fittings and the true 3D density profiles. The top and second rows are for the 2D elliptical fittings with FoV $14''\times14''$ and $6''\times6''$, respectively. The bottom panels are for the 2D radial fittings. The gray lines in each panel display the relative deviations for 200 randomly selected galaxies, while the red and blue lines with error bars show, respectively, the medians of the deviations for all 1362 elliptical galaxies and all 3981 disk galaxies. The error bars indicate the range of the first and third percentiles in each bin. The vertical dashed lines indicate the median of the half-light radii for all galaxies.}\label{fig:depro2d3d}
\end{figure*}

Figure \ref{fig:depro2d3d} presents the comparisons between the 3D density profiles (\ie~$\bar{\rho}_{\rm 2D}$, $\rho_{\rm 2D}$, and $j_{\rm 2D}$) predicted from 2D fittings and the true 3D density profiles. We can notice that the scatter is large for the deprojected profiles of the 2D fittings. We find that the inner mass density profiles for disk galaxies are significantly upturned due to the steeper inner slope of 2D fittings. However, for elliptical galaxies, the fitting bias is not significant in the inner region.

One may realize that a small negative bias exists around the $90\%$ light radius for the $\rho_{\rm 2D}$ predicted from 2D elliptical fittings with FoV $14''\times14''$ and 2D radial fittings. This is because the slopes around $r_{90}$ are relatively steeper for the 2D BPL fittings than for the corresponding 3D density profiles. However, this does not indicate that the mass density profiles are still underestimated at radius much larger than $r_{90}$. We examine the fitting bias at radius larger than two or three times $r_{90}$, \eg, and find that the true 3D mass density profiles are overestimated in general by the 2D fittings beyond a certain large radius.

\begin{figure*}[htb!]
  \centering
  \includegraphics[width=0.75\textwidth]{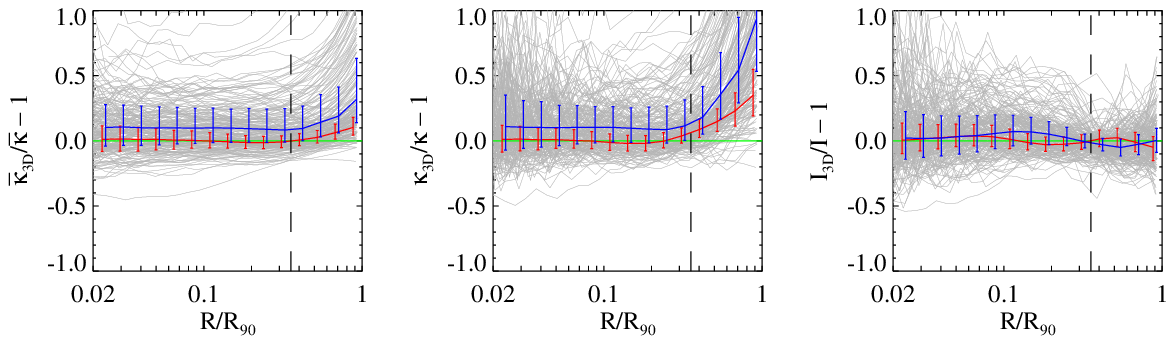}\\
  \caption{The relative differences between the surface density profiles (\ie~$\bar{\kappa}_{3D}$, $\kappa_{3D}$ and $I_{3D}$) by projecting the 3D radial fittings and the true surface density profiles. The gray and colored lines have similar meanings to those shown in Figure \ref{fig:depro2d3d}, but for radial surface density profiles.}\label{fig:pro3d2d}
\end{figure*}

In addition to the comparisons shown in Figure \ref{fig:depro2d3d}, we also compare the projections of 3D fittings (\ie~$\bar{\kappa}_{3D}$, $\kappa_{3D}$, and $I_{3D}$) with the true surface density profiles in Figure \ref{fig:pro3d2d}. It is noticeable that, for disk galaxies, the surface mass density profiles are significantly biased high from inner to outer regions by the projections of 3D BPL fittings, and the positive biases are more serious at larger radius. While, for elliptical galaxies, the true surface mass density profiles can be described very well by $\bar{\kappa}_{3D}$ and $\kappa_{3D}$ within the projected half-light radius, although overestimation still appears at larger radius.

The increasing trend of biases shown in Figure \ref{fig:pro3d2d} can be understood easily by considering the flatter outer slopes of 3D BPL model fittings. Because the true density profiles decrease faster than those estimated by BPL model fittings at larger radius, the relative biases are expected to be more significant at larger radius. Compared to disk galaxies, the biases for elliptical galaxies are systematically smaller because their outer slopes in 3D fittings are about $2$ and much steeper than the slopes ($\sim1.5$) for disk galaxies. Note that, as shown in Figure \ref{fig:3djoint}, the BPL model does not always overestimate the volume density profiles and may underestimate them within the fitting range. Therefore, by projection, the bias in the relatively inner region can be reduced and may be not significant, \eg, for elliptical galaxies.

In contrast to BPL model fittings, the last panel of Figure \ref{fig:pro3d2d} demonstrates the good performance of PL-S\'ersic model fittings to light density profiles.

In short, this subsection further illustrates the effects of projection, fitting ranges, and methods on density profile fittings. We compare in detail the true volume density profiles with the deprojections of the 2D fittings and the true surface density profiles with the projections of the 3D fittings. We find that the BPL model can give a reasonable prediction of the 3D density profiles within $r_{90}$ from the deprojections of the 2D fittings (see Figure \ref{fig:depro2d3d}). However, the BPL model has limitations. The 3D BPL fittings cannot be directly projected to estimate the surface mass density profiles, especially for disk galaxies (see Figure \ref{fig:pro3d2d}). As a lens mass model, the BPL model is mainly proposed to describe the surface mass distributions and the relevant lensing quantities. So, we care more about the accuracy of the 2D fittings and their consistency with the true volume density profiles within a certain radius. The large biases presented by the projections of the 3D BPL fittings are just for theoretical investigation and not likely to happen in actual lensing analyses.

\subsection{The Predicted AL-weighted LOSVDs}
Velocity dispersions can help us improve the constraints on the mass distributions of galaxies. In this subsection, we inspect whether the AL-weighted LOSVDs $\sigma_{\parallel,\rm pred}$ predicted from the BPL mass model fittings are consistent with the directly ``observed'' ones $\sigma_{\parallel,\rm sim}$.

\begin{figure*}[htb!]
  \centering
  \includegraphics[width=0.9\textwidth]{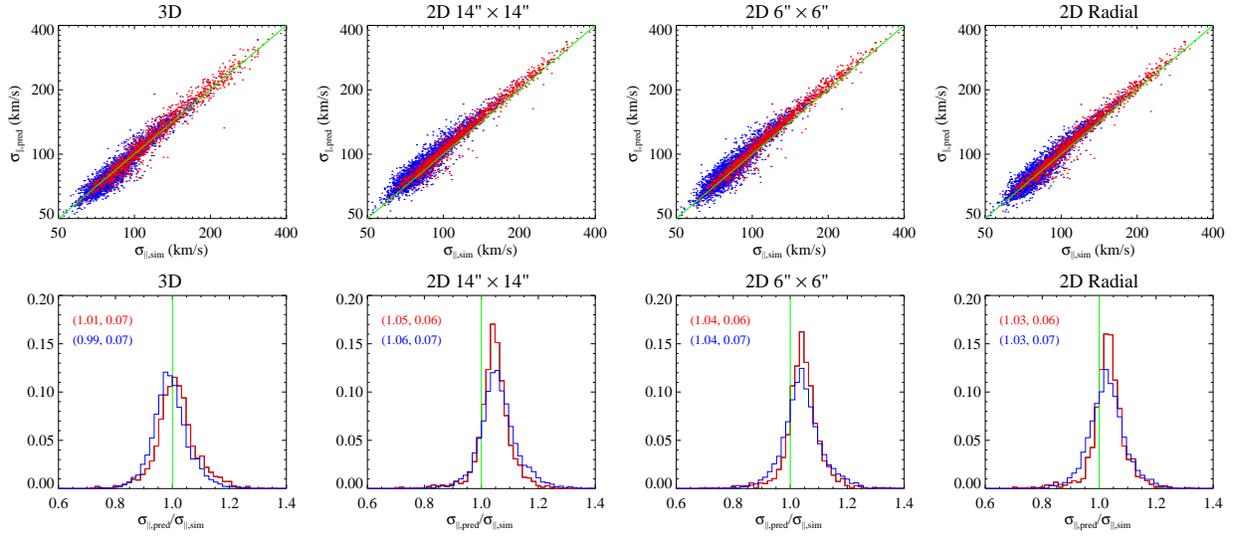}\\
  \caption{Comparisons between the ``predicted'' AL-weighted LOSVDs $\sigma_{\parallel,\rm pred}$ and the directly ``observed'' $\sigma_{\parallel,\rm sim}$. The first column shows the comparisons for the 3D fittings and the other columns are for the 2D fittings. The specific fitting method is indicated by the titles above each plot. The scatter diagrams display the one-to-one comparisons between $\sigma_{\parallel,\rm pred}$ and $\sigma_{\parallel,\rm sim}$. The histograms exhibit the distributions of the ratio $\sigma_{\parallel,\rm pred}/\sigma_{\parallel,\rm sim}$ for which the averages and standard deviations are presented using pairs of numbers. In these plots, the red and blue colors are for the elliptical and disk galaxies, respectively.}\label{fig:allosvd}
\end{figure*}

In Figure \ref{fig:allosvd}, we show the comparisons between the ``predicted'' $\sigma_{\parallel,\rm pred}$ and the true ``observed'' $\sigma_{\parallel,\rm sim}$ for the 3D and 2D fittings. It is found that the predicted and true AL-weighted LOSVDs are strongly correlated. For the 3D fittings, the bias is negligible, and the scatter is about $7\%$. However, for the 2D fittings, small positive biases exist as indicated by the positive shift of the scatter diagrams and the statistical distributions of $\sigma_{\parallel,\rm pred}/\sigma_{\parallel,\rm sim}$. The biases for the 2D fittings are larger than $3\%$ but typically less than $6\%$.

We find the the velocity dispersion bias can be corrected by accounting for the imperfect fittings of the BPL model and the projection effect. The BPL model can fit well the 2D mass distributions. However, the deprojection of the 2D fittings has significant scatter compared to the true 3D density profiles. The projection effect can complicate not only the reconstruction of mass distributions but also the prediction of AL-weighted LOSVDs.

For the 3D fittings, we conclude that there is no velocity dispersion bias (\ie, $b_\sigma = 1$) for the BPL model. For the 2D fittings, we find that the velocity dispersion bias can be roughly estimated by
\begin{equation}
  b_\sigma \simeq 1.015q_*^{-0.07}
\end{equation}
where $q_*(<1)$ is the axial ratio of the stellar mass distribution. The quantity $q_*$ can be estimated by the 2D elliptical PL-S\'ersic profile fitting or the inertia tensor of stellar particles within a certain area, \eg~ enclosed by the elliptical $90\%$ light radius $R_{\rm el,90}$. We find that the $b_\sigma$--$q_*$ relation is not very sensitive to the measuring methods of $q_*$ that we have investigated. The index $-0.07$ demonstrates the weak dependence of $b_\sigma$ on the observed ellipticity of galaxies. More spherical galaxies are less subject to the projection effect but still suffer from the limitation of mass models.

\begin{figure*}[htb!]
  \centering
  \includegraphics[width=0.9\textwidth]{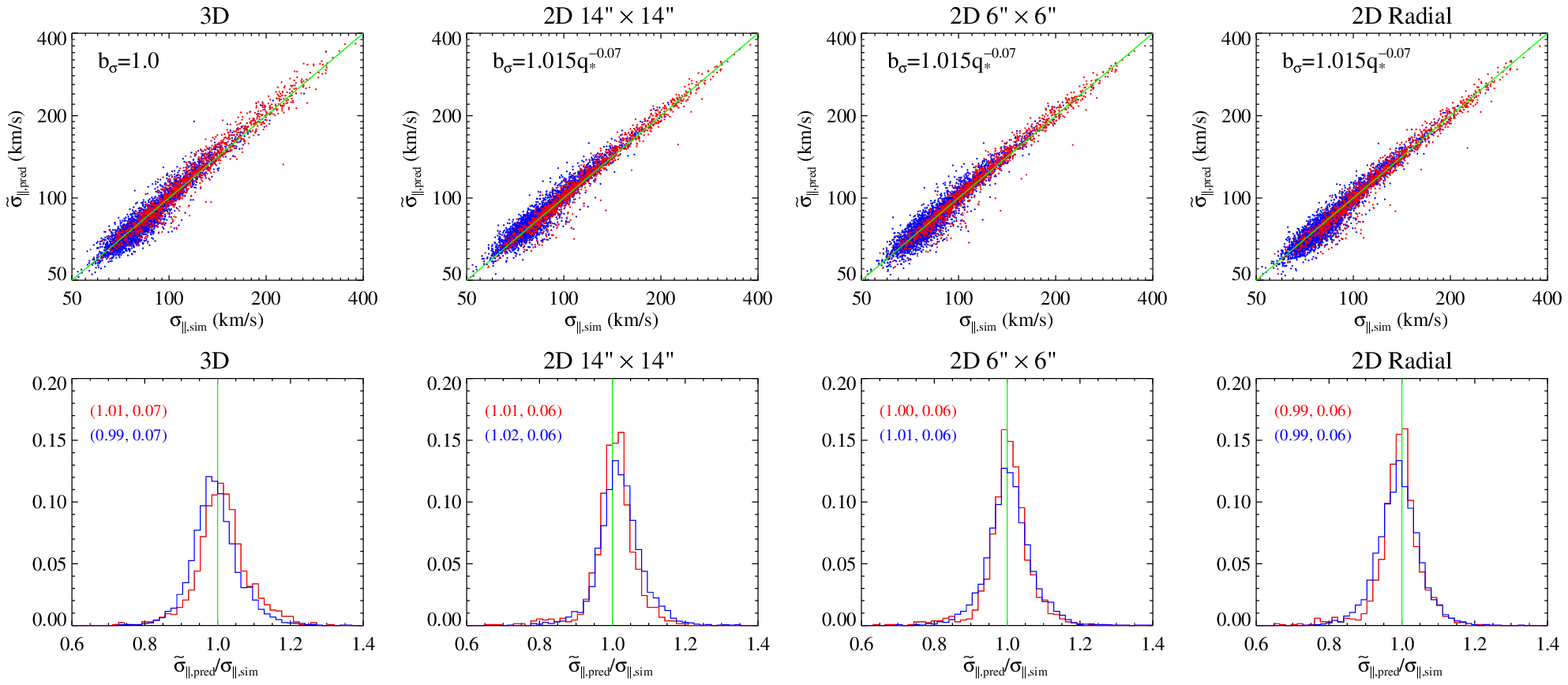}\\
  \caption{Comparisons between corrected $\tilde\sigma_{\parallel,\rm pred}=\sigma_{\parallel,\rm pred}/b_\sigma$ and $\sigma_{\parallel,\rm sim}$, where the $b_\sigma$ accounts for the bias due to BPL model fittings and the projection effect. Please refer to the caption of Figure \ref{fig:allosvd} for more details.}\label{fig:allosvdc}
\end{figure*}

By scaling the predicted $\sigma_{\parallel,\rm pred}$ by $b_\sigma$, we can then get the corrected AL-weighted LOSVD $\tilde\sigma_{\parallel,\rm pred}=\sigma_{\parallel,\rm pred}/b_\sigma$. In Figure \ref{fig:allosvdc}, we present the comparisons between $\tilde\sigma_{\parallel,\rm pred}$ and $\sigma_{\parallel,\rm sim}$. Evidently, the bias of $\sigma_{\parallel,\rm pred}$ from $\sigma_{\parallel,\rm sim}$ is well corrected for by the bias factor $b_\sigma$ for BPL model fittings. The velocity bias is finally reduced to be no more than $2\%$ for the 2D fittings, and the intrinsic scatter is typically about $6\%$.

\subsection{The Estimation of the Velocity Anisotropy Parameter}
It should be mentioned that the velocity anisotropy parameter $\beta$ used above for each galaxy is assumed to be a known quantity in the calculation of the AL-weighted LOSVD. However, in observations, $\beta$ cannot be easily determined even if the velocity dispersion profile is observed because of the degeneracy between the mass distribution and velocity anisotropy \citep{1993MNRAS.265..213G,2000A&AS..144...53K,2001MNRAS.322..702M,2010MNRAS.401.2433M}. By looking at the excellent consistency between $\tilde\sigma_{\parallel,\rm pred}$ and $\sigma_{\parallel,\rm sim}$ illustrated in Figure \ref{fig:allosvdc}, we propose to measure an effective velocity anisotropy $\beta_{\rm eff}$ by solving the equation $\sigma_{\parallel,\rm pred}(\beta)=b_\sigma \sigma_{\parallel,\rm sim}$.

\begin{figure*}[htb!]
  \centering
  \includegraphics[width=0.9\textwidth]{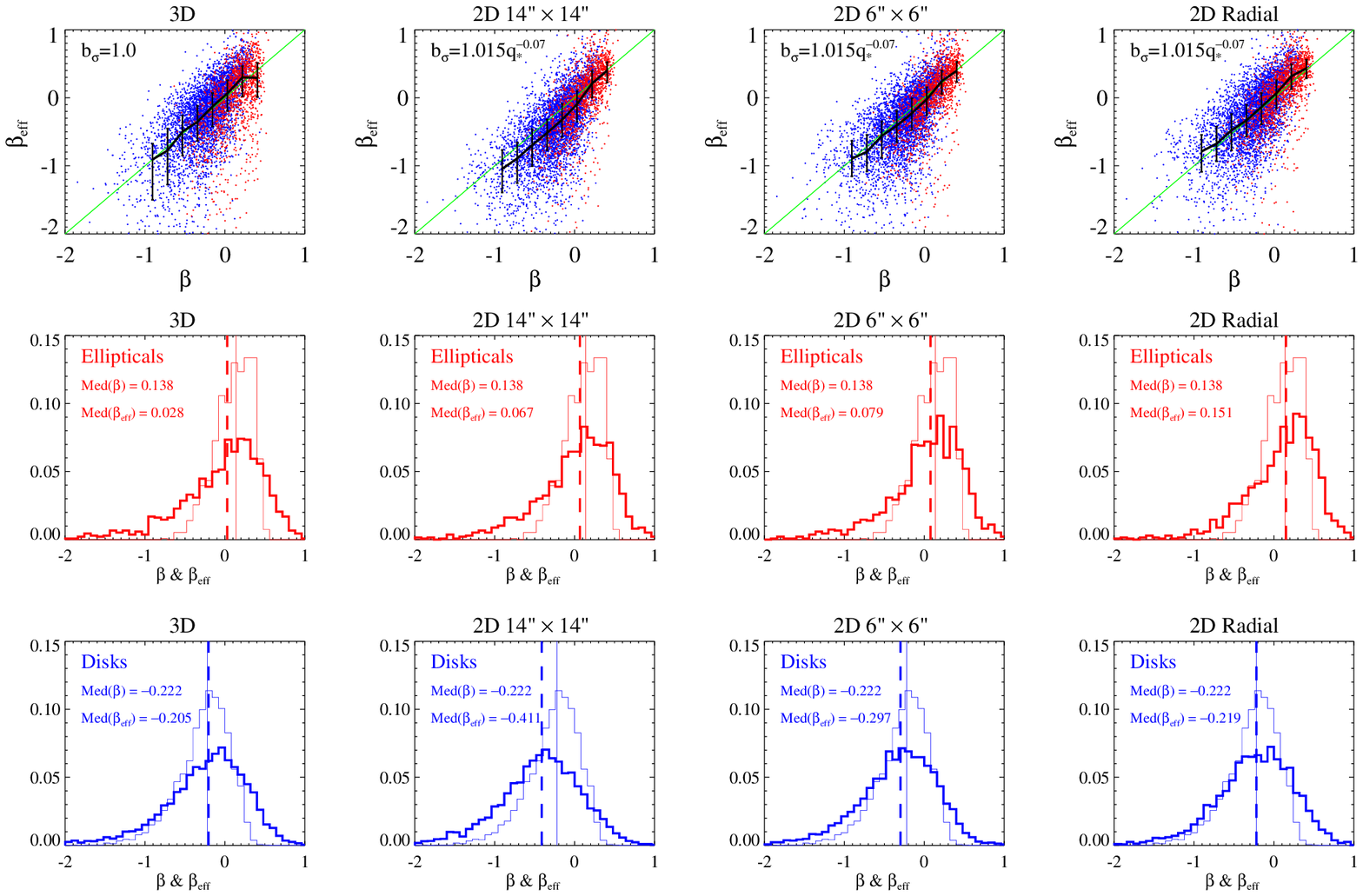}\\
  \caption{Comparisons between the predicted effective anisotropy $\beta_{\rm eff}$ and the directly measured $\beta$ in simulation. The red and blue dots in the top panels are for the elliptical and disk galaxies, respectively. The black lines present the median of $\beta_{\rm eff}$ as a function of $\beta$. The error bars indicate the range of the first and third percentiles in each bin. The second and third rows display the comparisons between the distributions of $\beta_{\rm eff}$ (thick histograms) and $\beta$ (thin histograms) for the elliptical (red) and disk (blue) galaxies, respectively. The thick dashed and thin solid vertical lines indicate the medians of the $\beta_{\rm eff}$ and $\beta$ distributions, respectively. The corresponding median values are presented in each panel.}\label{fig:pred_ani}
\end{figure*}

In Figure \ref{fig:pred_ani}, we show the comparisons between the predicted effective velocity anisotropy $\beta_{\rm eff}$ and the directly measured $\beta$ for all of the fitting methods. According to the scatter diagrams, we are aware that the uncertainties for the inferred $\beta_{\rm eff}$ are significant, demonstrating the difficulty in inferring the velocity anisotropy for an individual galaxy by resorting to the SL mass measurement and AL-weighted LOSVD observation. However, as indicated by the black lines, the median of the $\beta_{\rm eff}$ values for a sample of galaxies with nearly the same $\beta$ is basically unbiased for most of the fitting methods.

There exist negative biases (especially for the disk galaxies) for the 2D elliptical fittings with FoV $14''\times14''$. These biases may be corrected for by introducing a more general velocity bias factor $b_\sigma$, which also depends on the fitting area. However, we know that such a large FoV of $14''\times14''$ may not be necessary for galaxy-scale lensed image reconstructions in real observations.

We thus argue that it is possible to measure the distribution of velocity anisotropy parameters for certain types of galaxies. For instance, the second and bottom rows of Figure \ref{fig:pred_ani} present the comparisons between the distributions of $\beta_{\rm eff}$ (thick histograms) and $\beta$ (thin histograms) for the elliptical (red) and disk (blue) galaxies, respectively. We can see that the distributions of $\beta_{\rm eff}$ and $\beta$ are very well consistent with each other for most of the fittings. For the 2D fittings to the disk galaxies with FoV $14''\times14''$, an obvious negative bias exists for the predicted $\beta_{\rm eff}$ distributions because the velocity dispersion bias is not well corrected.

\section{Conclusion and Discussions}

In this paper, we propose the BPL profile as a lensing mass model to estimate the mass distributions of galaxies. The BPL model can be separated into a power-law part and a mass-complementary part in the central region. It can describe not only the mass distribution with a flat core but also that with an obvious cusp. More importantly, we find that the deflection angles and magnifications of the BPL model can be calculated analytically, making it an efficient lensing mass model.

The BPL model is validated by about $5000$ galaxies with stellar mass larger than $10^{10}h^{-1}\msun$ extracted from the Illustris-1 simulation. The 3D and 2D mass distributions of the simulated galaxies are fitted by the BPL profile. The corresponding light (or stellar mass) distributions are fitted by the PL-S\'ersic profile. Various fitting methods are considered, including a 3D and three 2D fittings. For all of the fitting methods, the mass and light density profiles are assumed to have the same break radius and inner density profile slope, in order to simplify the dynamical modeling.

As demonstrated by the 3D density profile fittings, the BPL model can well describe the volume mass density profiles of the Illustris galaxies within the $90\%$ light radius $r_{90}$. The inner mass density profiles can be distinct from the outer mass density profiles. The inner slopes for most galaxies are less than $1$ while the outer slopes are much steeper, \eg~ around $2$ for elliptical galaxies. We find that the biases of the 3D fittings are typically less than $10\%$ within $r_{90}$ for elliptical galaxies and somewhat larger for disk galaxies. Regardless of fitting methods, an overestimation trend always exists at quite a large radius for the BPL fittings. We find that the PL-S\'ersic profile is good enough to describe the light distributions.

For the 2D fittings, both the BPL mass and PL-S\'ersic light profile fittings perform well, especially for the elliptical galaxies. However, we find that the performance of the fittings is sensitive to the fitting area and procedures. For example, the fitted slopes are relatively flatter for smaller FoV. The 2D radial fittings with logarithmic radial bins can balance the fittings in the inner and outer regions. In any case, similar to the 3D fittings, the BPL model can overestimate the surface mass density profiles outside a certain radius.

We also inspect the consistency between the 2D and 3D BPL fittings. Because the true volume mass density profiles decrease faster at larger radius, the projection effect can make the 2D fitted slopes steeper than those of the 3D fittings, and slightly underestimate the volume mass density $\rho_c$ with a larger break radius $r_c$. By looking into the comparisons of true volume density profiles with the deprojections of 2D fittings, we find that, for elliptical galaxies, both the 3D mass and light density profiles can be recovered very well within $r_{90}$ by the 2D fittings. For disk galaxies, the deprojections of 2D BPL fittings are also workable, although significant biases may exist in the very central region.

In addition to the profile fittings to the mass distributions of simulated galaxies, we also study the BPL fittings to the more general density profiles, \eg~ the NFW and Einasto profiles, in the Appendix \ref{appendix:bplnfw}. We find that the BPL profile can mimic the 2D NFW and Einasto profiles within a certain radius. The deprojections of the 2D BPL fittings are also very well consistent with the true 3D density profiles, except for the Einasto density profiles with a too small Einasto index.

The BPL density profile fittings investigated in this paper prove that the BPL model is a more realistic lensing mass model. Although it tends to overestimate the mass distributions in the outer region of galaxies, it performs well within a sufficiently large radius, especially for the elliptical galaxies. We know that it is in practice impossible to constrain the mass distributions far away from the central region using solely SL observations, because SL images mainly provide information about the galaxies in the relatively central region. Thus, as a lensing mass model, the BPL model is good enough to be used for SL analyses.

Based on the BPL mass and PL-S\'ersic light density profiles, the AL-weighted LOSVDs are inspected mathematically and also using simulated galaxies. For the Illustris galaxies, we find that the predicted AL-weighted LOSVDs are correlated with the true values very well with only a small positive bias. This bias can be corrected for by accounting for the limitation of the BPL model and the projection effect. Realizing the strong correlation between the predicted AL-weighted LOSVDs and the true values, we propose a method to measure the distribution of velocity anisotropy parameters for a sample of galaxies that have SL and single-fiber spectroscopic observations. It is demonstrated that the distribution of velocity anisotropy parameters for a sample of galaxies can be well recovered.

To summarize, we investigate in this paper the basic properties of the BPL model and conclude that the BPL model is a more efficient and realistic lensing mass model for galactic and cosmological applications.

\acknowledgments
We thank the anonymous referee for comments and suggestions. We thank Kazuya Koyama and Dandan Xu for discussions. W.D. acknowledges the support from the National Natural Science Foundation of China (NSFC) under grants 11803043 and 11890691. This project is partly supported by the National Key Basic Research and Development Program of China (No. 2018YFA0404503 to G.B.Z. and No. 2018YFA0404501 to S.M.). We also acknowledge the support by NSFC grant Nos. 11925303, 11720101004, and 11673025 for G.B.Z.; Nos. 11333001 and 11933002 for Z.H.F.; Nos. 11821303, 11761131004, and 11761141012 for S.M. G.B.Z. and Z.H.F. are also supported by a grant of the CAS Interdisciplinary Innovation Team. Y.S. is supported by the Royal Society -- K.C. Wong International Fellowship (NF170995).

\appendix
\renewcommand\thefigure{\thesection.\arabic{figure}}
\renewcommand\theequation{\thesection.\arabic{equation}}
\section{The derivation of the analytical deflection angles for the BPL model}
\setcounter{figure}{0}
The analytical form of the deflection field can significantly speed up the SL analyses. We now show the derivation of the deflection angles of the BPL model in a little more detail. By Taylor-expanding the integrand of Equation (\ref{eq:alp_star}), we have
\begin{eqnarray}\label{eq:aseries}
\alpha^*(z)&=&\frac{2}{z}\sum_{n=0}^{\infty}\left(\frac{1}{2}\right)^{(n)}\frac{\zeta^{2n}}{n!}\int_0^{R_{\rm el}}\kappa(R)R^{2n+1} dR 
\end{eqnarray}
where $\zeta^2=(1/q-q)/z^2$ and $x^{(n)}={\Gamma(x+n)}/{\Gamma(x)}$ denotes the rising factorial of $x$. This series does not always converge except for $q\ge\sqrt{2}/2$. However, we note that this is not a problem for our aim of finding the analytical form of deflection angles.

Substituting the power-law part $\kappa_1(R)$ of the BPL model into the above equation, we then find the analytical expression of the deflection angle
\begin{eqnarray}\label{eq:alpkap1}
\alpha_1^*(z)&=&\frac{2}{z}\frac{3-\alpha}{2}b^{\alpha-1}\sum_{n=0}^{\infty}\left(\frac{1}{2}\right)^{(n)}\frac{\zeta^{2n}}{n!}\frac{R_{\rm el}^{2n+3-\alpha}}{2n+3-\alpha} \nonumber\\
&=& \frac{R_{\rm el}^2}{z}\left(\frac{b}{R_{\rm el}}\right)^{\alpha-1}\sum_{n=0}^{\infty}\frac{(\frac{1}{2})^{(n)}(\frac{3-\alpha}{2})^{(n)}}{(\frac{5-\alpha}{2})^{(n)}}
\frac{(\zeta R_{\rm el})^{2n}}{n!}  \nonumber\\
&=&\frac{R_{\rm el}^2}{z}\left(\frac{b}{R_{\rm el}}\right)^{\alpha-1}F\left(\frac{1}{2},\frac{3-\alpha}{2};\frac{5-\alpha}{2};\zeta^2R_{\rm el}^2\right) .
\end{eqnarray}

For the mass deficit or surplus part $\kappa_2(R)$, using Equation (\ref{eq:aseries}), one gets
\begin{equation}\label{eq:alpkap2}
\begin{split}%
\alpha_2^*(z)= & \frac{2}{z}\frac{3-\alpha}{ \mathcal{B}(\alpha)}\left(\frac{b}{r_c}\right)^{\alpha-1}\sum_{n=0}^{\infty}(\frac{1}{2})^{(n)}\frac{\zeta^{2n}}{n!}\int_0^{R_{\rm el}}
\tilde{z}\left[F\left(\frac{\alpha_c}{2},1;\frac{3}{2};\tilde{z}^2\right)-F\left(\frac{\alpha}{2},1;\frac{3}{2};\tilde{z}^2\right)\right] R^{2n+1} dR
\end{split}
\end{equation}
where $\tilde{z}=\sqrt{1-R^2/r_c^2}$. In view of that,
\begin{equation}
\begin{split}%
 & \int_0^{R_{\rm el}} \tilde{z}F\left(\frac{a}{2},1;\frac{3}{2};\tilde{z}^2\right) R^{2n+1} dR \\
=& \frac{r_c^{2n+2}}{2}\int_{\tilde{z}_{\rm el}^2}^{1}\sqrt{x}F\left(\frac{a}{2},1;\frac{3}{2};x\right)(1-x)^ndx \\
=& \frac{r_c^{2n+2}}{2}\left[\int_{0}^{1}\sqrt{x}(1-x)^nF\left(\frac{a}{2},1;\frac{3}{2};x\right)dx-
\tilde{z}_{\rm el}^{3}\int_{0}^{1}\sqrt{x}(1-\tilde{z}_{\rm el}^2x)^nF\left(\frac{a}{2},1;\frac{3}{2};\tilde{z}_{\rm el}^2x\right)dx\right] \\
=& \frac{r_c^{2n+2}}{2}\left[\frac{2}{3-a}\frac{1^{(n)}(\frac{3-a}{2})^{(n)}}{(\frac{3}{2})^{(n)}(\frac{5-a}{2})^{(n)}}-\mathcal{I}(n,a,\tilde{z}_{\rm el})\right]
\end{split}
\end{equation}
where $\tilde{z}_{\rm el}=\sqrt{1-R_{\rm el}^2/r_c^2}$ and
\begin{equation}
\mathcal{I}(n,a,\tilde{z}_{\rm el})=2\sum_{k=0}^{n}\binom{n}{k}(-1)^k\frac{\tilde{z}_{\rm el}^{2k+3}}{2k+3} {_3F_2}\left(\frac{2k+3}{2},\frac{a}{2},1;\frac{2k+5}{2},\frac{3}{2};\tilde{z}_{\rm el}^2\right) ,
\end{equation}
and one then finds
\begin{equation}\label{eq:alpkap2a}
\begin{split}%
\alpha_2^*(z)= & \frac{r_c^2}{z}\frac{3-\alpha}
  {\mathcal{B}(\alpha)}\left(\frac{b}{r_c}\right)^{\alpha-1}\sum_{n=0}^{\infty}\left(\frac{1}{2}\right)^{(n)}\frac{(r_c\zeta)^{2n}}{n!}
  \left[\frac{2}{3-\alpha_c}\frac{1^{(n)}(\frac{3-\alpha_c}{2})^{(n)}}{(\frac{3}{2})^{(n)}(\frac{5-\alpha_c}{2})^{(n)}}-\mathcal{I}(n,\alpha_c,\tilde{z}_{\rm el}) \right.{}\\
  &\left.+\mathcal{I}(n,\alpha,\tilde{z}_{\rm el})
  -\frac{2}{3-\alpha}\frac{1^{(n)}(\frac{3-\alpha}{2})^{(n)}}{(\frac{3}{2})^{(n)}(\frac{5-\alpha}{2})^{(n)}}\right] \\
= & \frac{r_c^2}{z}\frac{3-\alpha}{\mathcal{B}(\alpha)}\left(\frac{b}{r_c}\right)^{\alpha-1}
\left\{\frac{2}{3-\alpha_c}{_3F_2}\left(\frac{3-\alpha_c}{2},\frac{1}{2},1;\frac{5-\alpha_c}{2},\frac{3}{2};\mathcal{C}\right)-\right. \\
  &\left. \frac{2}{3-\alpha}{_3F_2}\left(\frac{3-\alpha}{2},\frac{1}{2},1;\frac{5-\alpha}{2},\frac{3}{2};\mathcal{C}\right)-
 \sum_{n=0}^{\infty}\left(\frac{1}{2}\right)^{(n)}\frac{\mathcal{C}^{n}}{n!}\left[\mathcal{I}(n,\alpha_c,\tilde{z}_{\rm el})-\mathcal{I}(n,\alpha,\tilde{z}_{\rm el})\right]\right\}
\end{split}
\end{equation}
where $\mathcal{C}=r_c^2\zeta^2=\frac{1-q^2}{q}\frac{r_c^2}{z^2}$.
For $R_{\rm el}>r_c$, the last series term disappears, giving the analytical expression in the region outside the core. However, within the break radius, the last term within the braces in Equation (\ref{eq:alpkap2a}) is very complicated and cannot be estimated efficiently. In the following, we will investigate the simplification of this series.

Using the power series representation of the Gauss hypergeometric function $F()$, the $\kappa_2(R)$ part of the BPL model, \ie~Equation (\ref{eq:kappa2}), can thus be rewritten as
\begin{eqnarray}
&&\kappa_2(R)=\frac{3-\alpha}{ \mathcal{B}(\alpha)}\left(\frac{b}{r_c}\right)^{\alpha-1}\times\left\{
           \begin{array}{ll}
             \displaystyle \left[\sum_{n=0}^{\infty}\frac{(\frac{\alpha_c}{2})^{(n)}}{(\frac{3}{2})^{(n)}}\tilde{z}^{2n+1}-
             \sum_{n=0}^{\infty}\frac{(\frac{\alpha}{2})^{(n)}}{(\frac{3}{2})^{(n)}}\tilde{z}^{2n+1}\right] &
\hbox{if $R\le r_c$} \\\\
             \displaystyle 0
& \hbox{if $R\ge r_c$ .}
           \end{array}
         \right. 
\end{eqnarray}
Inserting this equation into Equation (\ref{eq:alp_star}), we find that the corresponding deflection angle for $\kappa_2$ is
\begin{equation}\label{eq:alpkap2b}
\alpha_2^*(z)=\frac{r_{c}^2}{z}\frac{3-\alpha}{\mathcal{B}(\alpha)}\left(\frac{b}{r_c}\right)^{\alpha-1}
\left[\mathcal{S}(\alpha_c,\tilde{z}_{\rm el},\mathcal{C})-\mathcal{S}(\alpha,\tilde{z}_{\rm el},\mathcal{C})\right]
\end{equation}
with
\begin{equation}
\mathcal{S}(a,\tilde{z}_{\rm el},\mathcal{C})=
\frac{1}{\sqrt{1-\mathcal{C}}}\sum_{n=0}^{\infty}\frac{(\frac{a}{2})^{(n)}}{(\frac{3}{2})^{(n)}}\frac{2}{2n+3}
 \left[F\left(\frac{1}{2},\frac{2n+3}{2},\frac{2n+5}{2},\frac{\mathcal{C}}{\mathcal{C}-1}\right)-
\tilde{z}_{\rm el}^{2n+3}F\left(\frac{1}{2},\frac{2n+3}{2},\frac{2n+5}{2},\frac{\mathcal{C}\tilde{z}_{\rm el}^2}{\mathcal{C}-1}\right)\right] .
\end{equation}
Comparing Equation (\ref{eq:alpkap2b}) with Equation (\ref{eq:alpkap2a}), one finds
\begin{equation}\label{eq:alp2com1}
\frac{1}{\sqrt{1-\mathcal{C}}}\sum_{n=0}^{\infty}\frac{(\frac{a}{2})^{(n)}}{(\frac{3}{2})^{(n)}}\frac{2}{2n+3}
F\left(\frac{1}{2},\frac{2n+3}{2},\frac{2n+5}{2},\frac{\mathcal{C}}{\mathcal{C}-1}\right)
=\frac{2}{3-a}{_3F_2}\left(\frac{3-a}{2},\frac{1}{2},1;\frac{5-a}{2},\frac{3}{2};\mathcal{C}\right)
\end{equation}
and
\begin{equation}\label{eq:alp2com2}
\frac{1}{\sqrt{1-\mathcal{C}}}\sum_{n=0}^{\infty}\frac{(\frac{a}{2})^{(n)}}{(\frac{3}{2})^{(n)}}\frac{2\tilde{z}_{\rm el}^{2n+3}}{2n+3}
F\left(\frac{1}{2},\frac{2n+3}{2},\frac{2n+5}{2},\frac{\mathcal{C}\tilde{z}_{\rm el}^2}{\mathcal{C}-1}\right)
=\sum_{n=0}^{\infty}\left(\frac{1}{2}\right)^{(n)}\frac{\mathcal{C}^{n}}{n!}\mathcal{I}(n,a,\tilde{z}_{\rm el}) .
\end{equation}
Keeping the right side of Equation (\ref{eq:alp2com1}) and the left side of Equation (\ref{eq:alp2com2}), we then finally obtain the more efficient analytical deflection angle for the $\kappa_2(R)$,
\begin{equation}
\alpha_2^*(z)=\frac{r_c^2}{z}\frac{3-\alpha}{\mathcal{B}(\alpha)}\left(\frac{b}{r_c}\right)^{\alpha-1}
\left[\frac{2}{3-\alpha_c}\mathscr{F}\left(\frac{3-\alpha_c}{2},\mathcal{C}\right)-\frac{2}{3-\alpha}\mathscr{F}\left(\frac{3-\alpha}{2},\mathcal{C}\right)-
  \mathcal{S}_0(\alpha,\alpha_c,\tilde{z}_{\rm el},\mathcal{C})\right]
\end{equation}
where $\mathscr{F}$ and $\mathcal{S}_0$ have already been defined by the Equation (\ref{eq:hyp3f2s}) and (\ref{eq:series0}), respectively.

It is shown that the calculation of deflection angles for the BPL model strongly depends on the Gauss hypergeometric function $F()$, which is usually a built-in function in some programming languages. In this work, the accuracy and speed of $F()$ are inspected by the freely available function scipy.special.hyp2f1 in Python. We find that the $F()$ can be evaluated accurately in the whole complex plane by solving the divergence problem of function hyp2f1 in some regions \footnote{See the discussions in \url{https://github.com/scipy/scipy/pull/8151} and the improved source code scipy/special/specfun/specfun.f of function hyp2f1 in \url{https://github.com/scipy/scipy/pull/8151/files}}. We also find that, based on one of the quadratic transformations given by \citet{1881ASENS.2.10}, \ie~
\begin{equation}
\begin{split}
F\left(a,b,a+b+\frac{1}{2},x\right)=\left(\frac{2}{\sqrt{1-x}+1}\right)^{2a}F\left(2a,a-b+\frac{1}{2},a+b+\frac{1}{2},\frac{\sqrt{1-x}-1}{\sqrt{1-x}+1}\right) ,
\end{split}
\end{equation}
the Gauss hypergeometric function $F\left(\frac{1}{2},\frac{2n+3}{2},\frac{2n+5}{2},\frac{\mathcal{C}\tilde{z}_{\rm el}^2}{\mathcal{C}-1}\right)$ in  $\mathcal{S}_0$ can converge much faster. It is found that, in general, the computational speed of our analytic model to calculate the deflection angles is at least $10^3$ times faster than the numerical integrations in the core region and $10^4$ times faster outside the break radius.

\section{Examples of the effects of lens and source properties on lensed images}\label{appendix:lenses}
\begin{figure*}[htb!]
  \centering
  \includegraphics[width=0.9\textwidth]{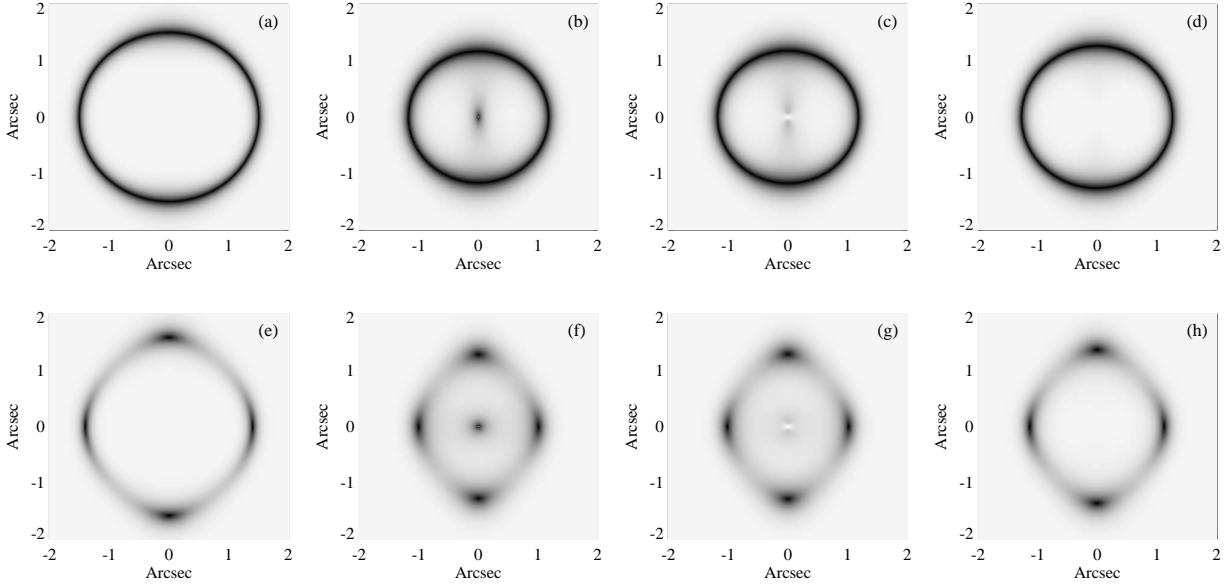}\\
  \caption{Lensed images corresponding to the lenses investigated in Figure \ref{fig:cri_cau}. The background source follows a S\'ersic profile with effective radius $R_{\rm eff}=0\farcs15$, S\'ersic index $n=1.0$, and ellipticity $q=0.5$. The source position is at $(0'',0'')$ and its major axis is along the $y$-direction.}\label{fig:arcs0}
\end{figure*}

\begin{figure*}[htb!]
  \centering
  \includegraphics[width=0.9\textwidth]{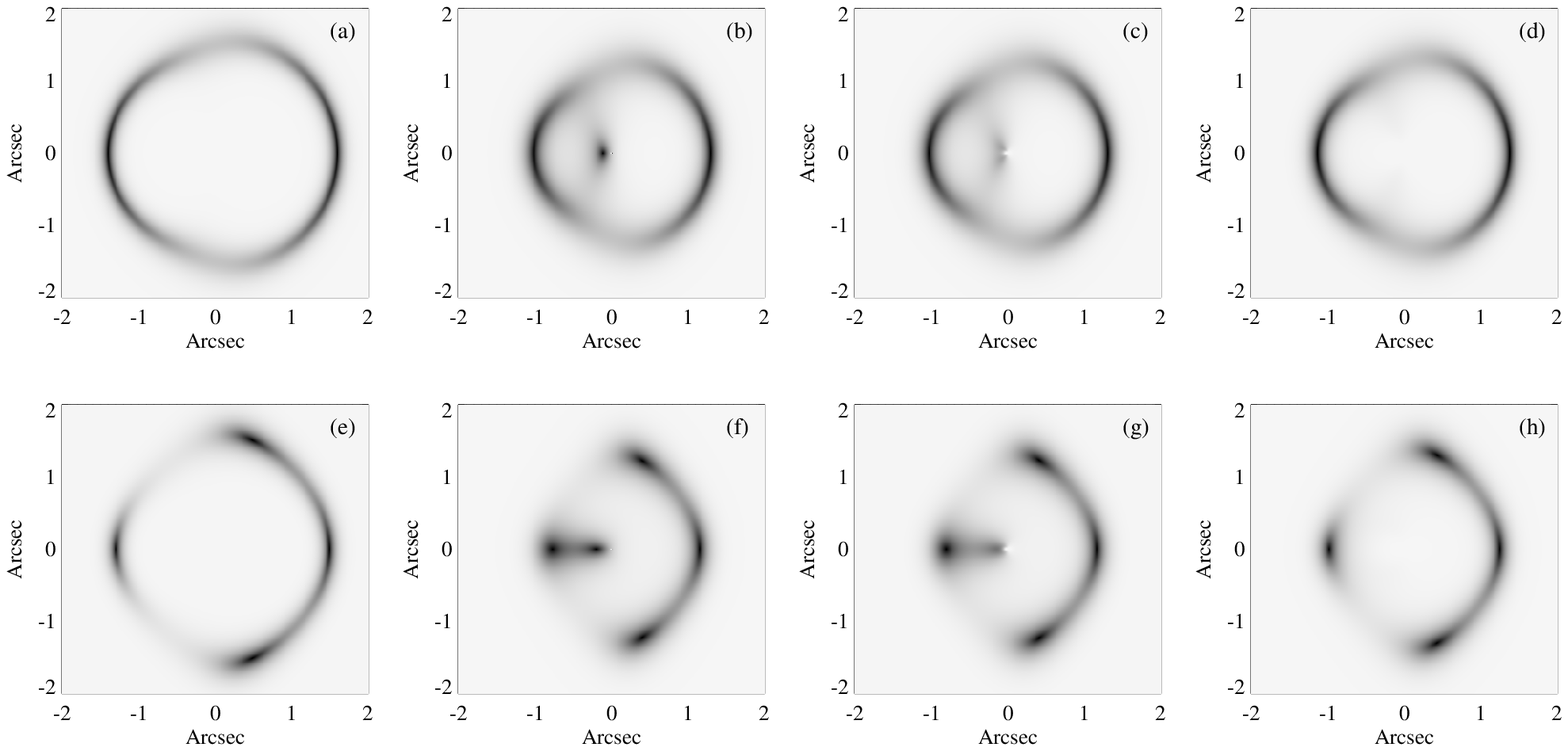}\\
  \caption{Similar to Figure \ref{fig:arcs0} but with source position $(0\farcs1,0)$.}\label{fig:arcs1}
\end{figure*}

\begin{figure*}[htb!]
  \centering
  \includegraphics[width=0.9\textwidth]{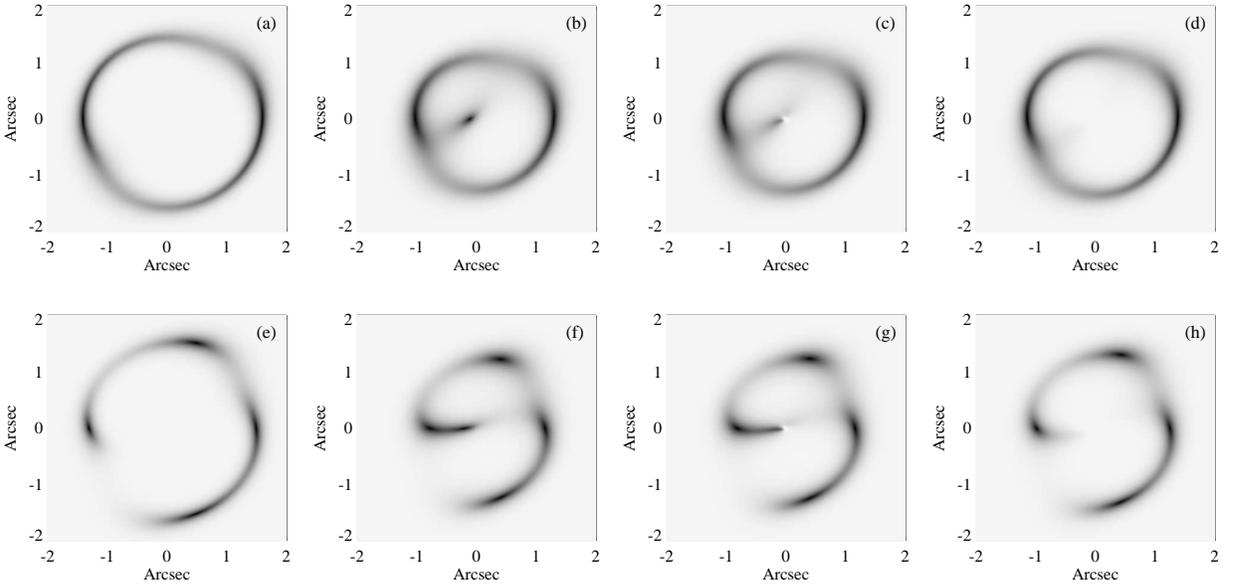}\\
  \caption{Similar to Figure \ref{fig:arcs0} but with the source centered at $(0\farcs1,0)$ and oriented along the $45^\circ$ direction.}\label{fig:arcs2}
\end{figure*}

Figures \ref{fig:arcs0}--\ref{fig:arcs2} illustrate the lensed images corresponding to the lenses investigated in Figure \ref{fig:cri_cau}. The background source follows a S\'ersic profile with effective radius $R_{\rm eff}=0\farcs15$, S\'ersic index $n=1.0$, and ellipticity $q=0.5$. The only differences of the sources in these three figures are the source positions and orientations. In Figure \ref{fig:arcs0}, the source is at the projected center of the lens and its major axis is along the $y$-direction. In Figure \ref{fig:arcs1}, the source is shifted to position $(0\farcs1,0)$ with the same orientation as that in Figure \ref{fig:arcs0}. However, in Figure \ref{fig:arcs2}, the source is at $(0\farcs1,0)$ and the position angle of its major axis is $45^\circ$.

As shown in these figures, the lensed image patterns around the critical curves are very similar to each other for the cases with the same lens ellipticity and source property. These results indicate that similar image configurations could be formed by different mass distributions.
It is also shown that the nonspherical shape of lenses can easily break the Einstein ring or giant arcs into multiple images. If the lens has a large flat core, there will be a central image or an image pattern extended to the center. However, a massive black hole will weaken the central image or make it disappear. Thus, it will be a challenge to constrain accurately the inner density profile of a lens by the lensed image itself if the central image is not evident.

On the other hand, lensed image patterns are also very susceptible to source properties. In real observations, due to the lack of knowledge about the intrinsic source properties, the so-called ``SPT'' can bring about more uncertainties to the SL-related analyses.

\section{BPL fittings to the NFW and Einasto profiles}\label{appendix:bplnfw}
In this Appendix, we further demonstrate the usability of the BPL model by looking into the BPL fittings to the more general density profiles, which, however, do not have the analytical form of deflection angles when the mass distribution is elliptically symmetric. The two-parameter NFW profile and the three-parameter Einasto profile are investigated here.

The NFW profile has the 3D form written as
\setcounter{equation}{0}
\begin{equation}\label{eq:rho_nfw}
\rho_{\scriptscriptstyle N}(r)=\frac{\rho_s}{(r/r_s)(1+r/r_s)^2},
\end{equation}
where $\rho_s$ is the characteristic density and $r_s=r_{-2}$ is the scale radius where the logarithmic slope of the density profile is $-2$. The corresponding total mass within $r$ is
\begin{equation}\label{eq:m_nfw}
M(r)=4\pi\rho_s r_s^3 \left[\ln(1+r/r_s)-r/(r+r_s)\right] .
\end{equation}

In practice, the $\rho_s$ and $r_s$ are usually replaced by the total mass $\mdlt$ within a radius $\rdlt$ and the concentration $c=\rdlt/r_s$, where $\rdlt$ denotes the radius within which the mean density is $\Delta$ times the critical density $\rho_{\rm crit}$ of the universe \citep{2002ApJ...574..538J,2008MNRAS.387..536G,2014ApJ...785...57D}. Given the mass $\mdlt$ and concentration $c$, we can immediately find $\rho_s=\delta_c\rho_{\rm crit}$ and $r_s=\rdlt/c$, where
\begin{equation}
\delta_c=\frac{\Delta}{3}\frac{c^3}{\ln(1+c)-c/(1+c)} .
\end{equation}

For the Einasto profile, its volume density profile has the same form as the 2D S\'ersic profile \citep{1965TrAlm...5...87E,2005ApJ...624L..85M,2012A&A...540A..70R}. We express the Einasto profile here as
\begin{equation}
\rho_{\scriptscriptstyle E}(r)=\rho_0\exp\left[-2n\left(\frac{r}{r_s}\right)^{\frac{1}{n}}\right] ,
\end{equation}
where $n$ is named as the Einasto index (analogous to the S\'ersic index), $\rho_0$ is the central density, and $r_s=r_{-2}$ is the scale radius where the logarithmic slope is $-2$. The total mass within $r$ is
\begin{equation}\label{eq:m_ein}
M(r)=4\pi\rho_0 r_s^3 \frac{n\gamma\left[3n,2n(\frac{r}{r_s})^{\frac{1}{n}}\right]}{(2n)^{3n}} ,
\end{equation}
where $\gamma(a,x)$ is the lower incomplete gamma function. Similar to the NFW profile, we can also introduce the mass $\mdlt=4\pi r_{\scriptscriptstyle\Delta}^3\Delta\rho_{\rm crit}/3$ and concentration $c=\rdlt/r_s$, and define $\rho_0=\delta_c\rho_{\rm crit}$ with
\begin{equation}
\delta_c=\frac{\Delta}{3}\frac{c^3(2n)^{3n}}{n\gamma(3n,2nc^{\frac{1}{n}})} .
\end{equation}
The surface density profiles $\Sigma_{\scriptscriptstyle N}(R)$ for the NFW profile \citep[\eg][]{2000ApJ...534...34W} and $\Sigma_{\scriptscriptstyle E}(R)$ for the Einasto profile \citep[\eg][]{2010MNRAS.405..340D,2012A&A...540A..70R} can be obtained, respectively, by integrating the volume density profiles $\rho_{\scriptscriptstyle N}(r)$ and $\rho_{\scriptscriptstyle E}(r)$ along the line of sight.

In the following analyses, we concentrate on the BPL fittings to the surface mass density profiles of the NFW and Einasto models. In accordance with the lensing observations, the surface mass density profiles $\Sigma_{\scriptscriptstyle N}(R)$ and $\Sigma_{\scriptscriptstyle E}(R)$ are scaled by the surface critical density $\Sigma_{\rm crit}=4.0\times10^{15}\msun/\mpc^{2}$ for a lens system with lens redshift $z_d=0.178$ and source redshift $z_s=0.6$. We also compare the true 3D density profiles with those predicted from 2D BPL fittings.

Figure \ref{fig:bplfitn} shows the BPL fittings to four spherical NFW profiles with the same mass $M_{200}=10^{13}\msun$ defined by $\Delta=200$. Four concentrations are considered with $c=3$, $5$, $10$, and $20$. We present the fits to the radial convergence profiles in the top panels and the comparisons between the true and predicted 3D density profiles in the bottom panels. It is found that the 2D NFW profiles can be fitted very well by the BPL model in the region we are focusing on. The true 3D density profiles can also be well estimated by the deprojections of the 2D BPL fittings. Note that the BPL parameter values here are sensitive to the fitting range. For example, as the upper limit of the fitting range increases, all of the BPL parameter values ($b$, $\alpha$, $r_c$ and $\alpha_c$) tend to be larger.

Figure \ref{fig:bplfite} shows the BPL fittings to four cases of spherical Einasto profiles. The mass and concentrations of the Einasto profiles are the same as those of the corresponding NFW profiles investigated above. By inspecting the profile fittings, we can notice that the 2D BPL fittings are also quite good for the Einasto density profiles within the fitting range. However, as shown in the bottom panels, large deviations may exist in the fitting range for the predicted 3D BPL density profiles. For instance, the deviations in the very central region can be significant for the Einasto profiles with extremely large and flat cores. The reason is that the Einasto profiles are nonsingular but fitted with the singular BPL profiles. However, this may not be a problem for massive galaxies which may always be singular due to the existence of a central massive black hole.

\setcounter{figure}{0}
\begin{figure*}[htb!]
  \centering
  \includegraphics[width=0.9\textwidth]{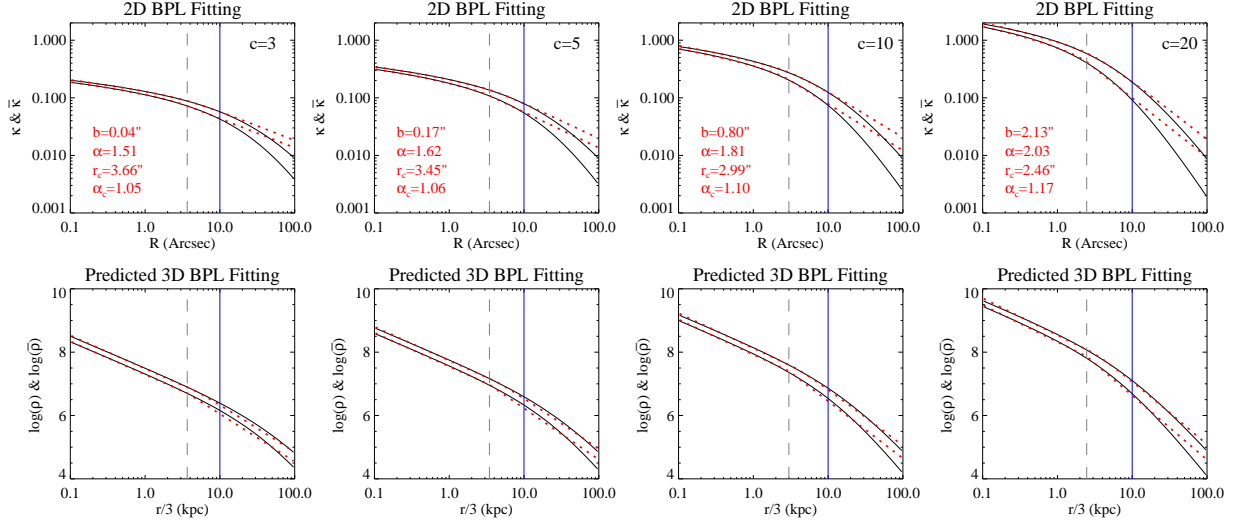}\\
  \caption{BPL fittings to four spherical NFW profiles with the same mass $M_{200}=10^{13}\msun$ and different concentrations $c=3$, $5$, $10$, and $20$. The top panels present the fits to the radial convergence profiles while the bottom panels show the comparisons between the true 3D NFW profiles and the predicted 3D profiles from 2D BPL fittings. In the top panels, the black lines show the mean convergence ($\bar{\kappa}$ with higher values) and convergence ($\kappa$ with lower values) profiles. The red dotted lines show the 2D BPL fittings to the black lines in the range of $0\farcs1$ to $10''$. The vertical blue lines mark the maximum fitting radius of $10''$. The best-fit parameter values are displayed with red numbers. The vertical dashed line in each panel indicates the break radius $r_c$ of the BPL fitting. In the bottom panels, the black lines present the original 3D NFW density profiles $\rho(r)$ with lower values and the mean density profiles $\bar{\rho}(r)$ with higher values in units of $\msun/\kpc^3$. The red dotted lines are not the fittings to the true density profiles but the 3D density profiles predicted directly from the 2D BPL fittings shown in the top panels. Note that the radius $r$ in the bottom panels is scaled by $3$ in order to be consistent with the angular scale used in the top panels.}\label{fig:bplfitn}
\end{figure*}

\begin{figure*}[htb!]
  \centering
  \includegraphics[width=0.9\textwidth]{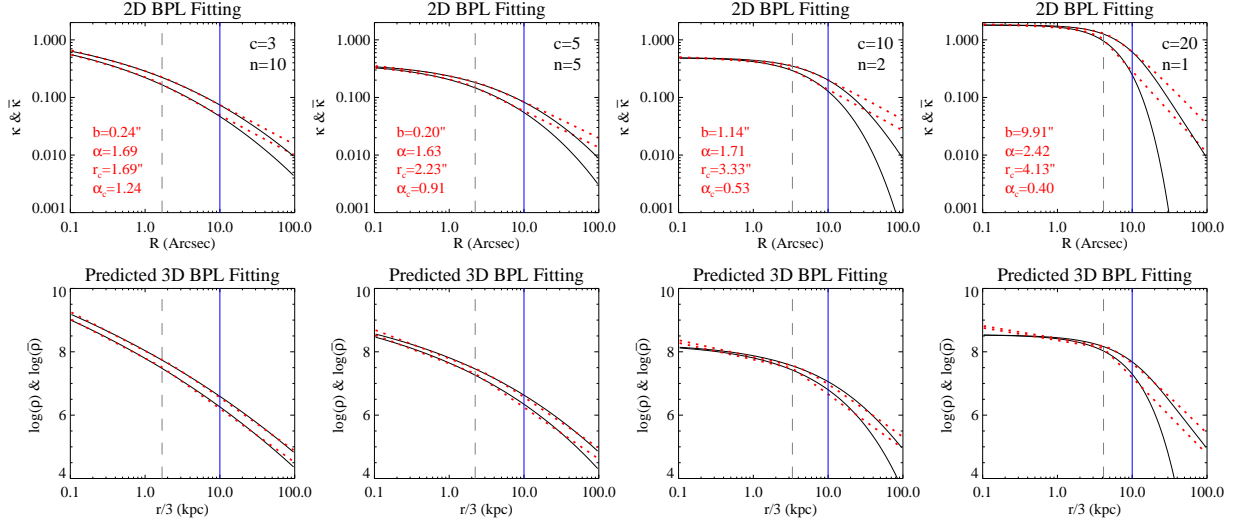}\\
  \caption{Similar to Figure \ref{fig:bplfitn} but for the BPL fittings to Einasto profiles with the same mass $M_{200}=10^{13}\msun$. The corresponding concentration $c$ and Einasto index $n$ are shown for each profile in the top panels.}\label{fig:bplfite}
\end{figure*}

\end{document}